\documentclass[11pt,preprint]{aastex}
\usepackage{natbib}
\usepackage{color}
\usepackage{graphicx}
\usepackage{rotating}
\usepackage{hyperref}
\usepackage[all]{hypcap}
\usepackage{mathrsfs}
\usepackage{amsmath}
\usepackage{amssymb}
\hypersetup{
    bookmarks=true,         
    unicode=false,          
    pdftoolbar=true,        
    pdfmenubar=true,        
    pdffitwindow=false,     
    pdfstartview={FitH},    
    pdftitle={My title},    
    pdfauthor={Author},     
    pdfsubject={Subject},   
    pdfcreator={Creator},   
    pdfproducer={Producer}, 
    pdfkeywords={keyword1} {key2} {key3}, 
    pdfnewwindow=true,      
    colorlinks=true,        
    linkcolor=blue,         
    citecolor=blue,         
    filecolor=magenta,      
    urlcolor=cyan           
}

\begin{document}
\title{Origin and Structures of Solar Eruptions II: Magnetic Modeling \\
(Invited Review)}
\author{Yang Guo$^{1,2}$, Xin Cheng$^{1,2}$, M. D. Ding$^{1,2}$}

\affil{$^1$ School of Astronomy and Space Science, Nanjing University, Nanjing 210023, China} \email{guoyang@nju.edu.cn}
\affil{$^2$ Key Laboratory for Modern Astronomy and Astrophysics (Nanjing University), Ministry of Education, Nanjing 210023, China}

\begin{abstract}
The topology and dynamics of the three-dimensional magnetic field in the solar atmosphere govern various solar eruptive phenomena and activities, such as flares, coronal mass ejections, and filaments/prominences. We have to observe and model the vector magnetic field to understand the structures and physical mechanisms of these solar activities. Vector magnetic fields on the photosphere are routinely observed via the polarized light, and inferred with the inversion of Stokes profiles. To analyze these vector magnetic fields, we need first to remove the 180$^\circ$ ambiguity of the transverse components and correct the projection effect. Then, the vector magnetic field can be served as the boundary conditions for a force-free field modeling after a proper preprocessing. The photospheric velocity field can also be derived from a time sequence of vector magnetic fields. Three-dimensional magnetic field could be derived and studied with theoretical force-free field models, numerical nonlinear force-free field models, magnetohydrostatic models, and magnetohydrodynamic models. Magnetic energy can be computed with three-dimensional magnetic field models or a time series of vector magnetic field. The magnetic topology is analyzed by pinpointing the positions of magnetic null points, bald patches, and quasi-separatrix layers. As a well conserved physical quantity, magnetic helicity can be computed with various methods, such as the finite volume method, discrete flux tube method, and helicity flux integration method. This quantity serves as a promising parameter characterizing the activity level of solar active regions.
\end{abstract}


\section{Introduction}
To understand the origin and structures of solar activities and the related phenomena, such as flares, filaments/prominencs, and coronal mass ejections (CMEs), we have to know the three-dimensional (3D) solar magnetic field, since the solar atmosphere is filled with magnetized plasma. Due to the high conductivity in the solar atmosphere, the magnetic field is frozen to the plasma. From the upper chromosphere to the transition region and the lower corona, the magnetic pressure even dominates over the gas pressure of the plasma. Therefore, the topology and dynamics of the magnetic field are critical to controlling the structure and behavior of plasma in the solar atmosphere.

Magnetic structures before eruptions could be either a magnetic flux rope, or sheared magnetic arcades, or both. A magnetic flux rope is defined as a bundle of magnetic field lines twisting around a common axis. Sheared arcades are regarded as a bundle of magnetic field lines that deviates far from the potential state but does not possess an inverse polarity, that is, a magnetic field component pointing from the negative to the positive polarity. Namely, sheared arcades possess a moderate twist (probably less than one turn) and electric current, but they are not strong enough to generate the inverse polarity. A magnetic flux rope can be formed from sheared arcades by magnetic reconnection and footpoint twisting motion.

The central physical mechanisms for solar activities include magnetohydrodynamic (MHD) instabilities and magnetic reconnection. For example, magnetic flux rope eruptions are explained by the loss of equilibrium \citep{1988Demoulin,1991Forbes,2000Lin2} or torus instability \citep{1978Bateman,2006Kliem}. \citet{2010Demoulin} pointed out that these two ideas are two different views of the same mechanism, namely, they both resort to the Lorentz repulsion force. One could increase the electric current in a magnetic flux rope or decrease the background magnetic field strength to build the state that leads to loss of equilibrium or the torus instability, either via an ideal process (e.g., kink instability) or a resistive process (magnetic reconnection). The magnetic reconnection could occur either under a flux rope or sheared arcades in the tether cutting model \citep{1980Moore,2001Moore}, or, above sheared arcades in the breakout model \citep{1998Antiochos,1999Antiochos}. Magnetic field observations and modeling are critical to understand how these physical mechanisms interplay with each other in specific events.

Studies on the magnetic field can be performed through observations, theoretical models, and numerical models. In this review, we first introduce the vector magnetic field observations and processing in Section~\ref{sec:vector}. The observations of Stokes profiles from the polarized lights in the solar atmosphere are mentioned. The magnetic field information is extracted from the inversion of Stokes profiles (Section~\ref{sec:inversion}). For vector magnetic field, the transverse components have an intrinsic 180$^\circ$ ambiguity, which has to be removed before further analysis (Section~\ref{sec:ambiguity}). The correction of the projection effect is also very important for solar vector magnetic field, especially when the region of interest is close to the solar limb or the field of view is large (Section~\ref{sec:projection}). Velocities can be derived from a time series of vector magnetic field by the optical flow techniques (Section~\ref{sec:velocity}).

At present, magnetic field in the solar atmosphere is observed routinely and relatively accurately only on the photosphere. The 3D magnetic field above the photosphere could be studied purely theoretically, or constructed numerically from observations. In Section~\ref{sec:extrapolation}, we introduce some theoretical force-free field models (Section~\ref{sec:theory}), numerical nonlinear force-free field models (Section~\ref{sec:nlfff}), magneto-hydrostatic (MHS) models (Section~\ref{sec:mhs}), and MHD models (Section~\ref{sec:mhd}) associated with solar magnetic flux rope emergence or eruptions. A force-free field is defined as a magnetic field without any Lorentz force. Typical force-free field models include the potential field, linear force-free field, and nonlinear force-free field. If the vector magnetic field on the photosphere is used as the boundary condition for a nonlinear force-free field modeling, preprocessing of the vector magnetic field is also essential for removing the Lorentz force on the boundary (Section~\ref{sec:prepro}). Since force-free field models are static, they are not suitable for studying dynamic solar eruptive phenomena, which have been widely studied by MHD models.

Magnetic energy computation in a volume and from the boundary is briefly discussed in Section~\ref{sec:energy}. To study in detail what are the critical features in a magnetic field for MHD instability and magnetic reconnection, we have to know the magnetic topology. The magnetic topology analysis includes searching for the locations of magnetic null points (Section~\ref{sec:null}), bald patches (Section~\ref{sec:bald}), and quasi-separatrix layers (QSLs; Section~\ref{sec:qsl}), and analyzing the magnetic structures and evolutions in light of these specific topologies. We also introduce the applications of magnetic topology analysis to observations in Section~\ref{sec:application}. The helicity computation is discussed in Section~\ref{sec:helicity}, where we focus on the finite volume method (Section~\ref{sec:fv}), discrete flux tube method (Section~\ref{sec:dt}), and helicity flux integration method (Section~\ref{sec:fi}). Finally, we give a summary and discussion in Section~\ref{sec:conclusion}.

This paper is focused on the magnetic field observations and modelings of various solar eruptive activities. The multi-wavelength observations of the origin and structures of CMEs, flares, and magnetic flux ropes are presented in another review by \citet{2017Cheng}.

\section{Vector Magnetic Field Observations and Processing} \label{sec:vector}
In the photosphere with relatively high density and low temperature, the vector magnetic field is routinely observed by space telescopes, such as the Solar Optical Telescope \citep[SOT;][]{2008tsuneta,2008suematsu,2008ichimoto,2008shimizu} on board \textit{Hinode} \citep{2007Kosugi} and the Helioseismic and Magnetic Imager \citep[HMI;][]{2012Scherrer,2012Schou} on board the \textit{Solar Dynamics Observatory} (\textit{SDO}). There are also various ground-based telescopes aiming at observing solar magnetic field, such as the Solar Magnetic Field Telescope (SMFT) of Huairou Solar Observing Station of National Astronomical Observatory of China \citep{1987Ai,1994Zhang,2004Su,2007Su}, the Multi-Raies (MTR) mode of the T\'elescope H\'eliographique pour l'Etude du Magn\'etisme et des Instabilit\'es Solaires \citep[THEMIS;][]{2000LopezAriste,2007Bommier}, the New Solar Telescope \citep[NST;][]{2010Cao,2017Wang} at Big Bear Solar Observatory, and so on.

Magnetic field in the photosphere is measured by the polarized light generated by the Zeeman splitting \citep{1908Hale}. Inference of the vector magnetic field requires the knowledge of the spectral profile of the Stokes parameters, $I, Q, U,$ and $V$. There are basically two types of instruments to measure the Stokes profiles, the filter type and the spectrograph type. HMI and SMFT belong to the filter type. They cover a large field of view, full solar disk for HMI and $4' \times 6'$ for SMFT, have a fairly high spatial resolution, $0.5''$ per pixel for HMI and $0.4'' \times 0.7''$ for SMFT, and have a relatively high cadence, 12 minutes for HMI vector magnetic field and 1 minute for SMFT. However, their spectral resolution is low. For example, HMI measures the Stokes parameters at six wavelengths with a band width of 76 m\AA \ evenly sampled across the Fe I 6173~\AA \ line covering a tunable range of 690 m\AA , and SMFT acquires data at two wavelengths ($-75$ m\AA \ and the line center) with a band width of 125 m\AA \ across the Fe I 5324~\AA \ line.

The Spectropolarimeter (SP) of SOT and MTR of THEMIS belong to the spectrograph type. They have very high spectral resolutions. For example, SOT/SP has a spectral sample of 21.5~m\AA \ per pixel through the spectral range of 6300.8~\AA \ to 6303.2~\AA \ and THEMIS/MTR has a dispersion of 19.5~m\AA \ for the spectral lines 6301.5 and 6302.5~\AA . But their field of view is restricted to a slit, $164'' \times 0.16''$ for SOT/SP and $120'' \times 0.5''$ for THEMIS/MTR, respectively. SOT/SP spends 4.8 seconds for observation at one slit position and THEMIS/MTR spends about 3 seconds for each scan step. To cover a large field of view, they have to take tens of minutes to several hours to scan over the region of target by moving perpendicular to the slit direction.

\subsection{Inversion of Stokes Profiles} \label{sec:inversion}

To extract the vector magnetic field information from observed polarized spectra, we have to resort to the inversion of Stokes profiles, namely, the inversion of spectral lines of $I, Q, U,$ and $V$. It faces two problems, namely, the forward modeling problem and the inversion problem. In the forward modeling problem, one has to solve the radiative transfer equation of polarized radiation to study how the Stokes profiles are formed. Pioneered by \citet{1956Unno}, this problem has been further discussed by many other authors by considering the magneto-optic, damping, and other effects \citep{1962Rachkovsky,1982Landolfi,1989Jefferies}. This model includes three parameters for the vector magnetic field, namely the field strength $B$, inclination angle $\psi$, and the azimuth angle $\phi$, seven parameters for the thermodynamics, namely, the line strength $\eta_0$, Doppler width $\Delta \lambda_D$, damping constant $a$ or equivalently damping parameter $\gamma$ with $a=\gamma/(4\pi \Delta \lambda_D)$, Doppler velocity $v$ or equivalently line wavelength $\lambda_0$, source function constant $B_0$, source function gradient $B_1$, and macro-turbulent velocity $v_\mathrm{m}$, and finally one geometrical parameter, the filling factor $\alpha$. To simplify the solution of the radiative transfer equations of the Stokes profiles, all the parameters are assumed to be constant as a function of the continuum optical depth $\tau$, except the source function $B(\tau)$, which is assumed to be linearly dependent on $\tau$ with $B(\tau)=B_0+B_1 \tau$.

The inversion problem fits the modeled Stokes profiles to the observations by adjusting the aforementioned parameters. This is a nonlinear least square problem, which is usually solved by the Levenberg--Marquardt algorithm \citep{1963Marquardt,1988Press} to minimize the discrepancies, indicated by an objective function $\chi^2$, between the modeled and observed Stokes profiles. Main efforts are devoted to search for the global minimum of $\chi^2$ in the parameter space discussed above. There are already some inversion procedures based on the Unno--Rachkovsky solutions and the Levenberg--Marquardt algorithm, such as the Milne--Eddington Line Analysis using an Inversion Engine \citep[MELANIE;][]{2001Socas-Navarro}, the UNNOFIT inversion code \citep{1982Landolfi,2007Bommier}, and the Very Fast Inversion of the Stokes Vector \citep[VFISV;][]{2011Borrero,2014Centeno}. These procedures have been applied to the observations of the Stokes profiles by different instruments, for example, MELANIE to SOT/SP, UNNOFIT to THEMIS/MTR, and VFISV to HMI.

\citet{2010Guo3} presented a detailed comparison of MELANIE and UNNOFIT and found that both procedures provide consistent inversion results of the field strength $B$, inclination angle $\psi$, and the azimuth angle $\phi$. \citet{2014Teng} made some tests and improvements to VFISV. They found that using a smooth interpolation for the Voigt function could eliminate spurious inversion results, and using proper initial values for the azimuth angle $\phi$ could speed up the code by four times relative to the original one, but provide accurate results only for the vector magnetic field and Doppler velocity. \citet{2015Teng} applied a kernel-based machine learning method to the inversion of Stokes parameters observed by HMI, and a fast inversion method is further proposed by \citet{2016Teng} based on the quadratic regression.

\subsection{Removing the 180$^\circ$ Ambiguity of the Transverse Components} \label{sec:ambiguity}

The 180$^\circ$ ambiguity of the transverse components of a vector magnetic field arises from the intrinsic symmetry (or, periodicity) of the radiative transfer of polarized lights (see, e.g., Equation~(28) of \citealt{1989Jefferies}). If the azimuth of the vector magnetic field changes 180$^\circ$, the emergent Stokes profiles are identical to those with the original azimuth. Considering the physical nature of the Zeeman splitting and the mathematical models of the aforementioned Stokes profile formation, there is no known method to resolve this ambiguity. Although it has been proposed to measure the vector magnetic field at more than one heights to resolve the ambiguity using the solenoidal property, determination of the formation heights of different spectral lines turns out to be difficult \citep{2013Bommier}. Therefore, the ambiguity is usually resolved by additional physical assumptions on the magnetic field.

Different assumptions have been made to constrain the magnetic field. For example, the potential field model is adopted as an reference model in many algorithms (see \citealt{2006Metcalf}), where the ambiguity is resolved by assuming the observed components make an acute angle with the modeled ones. The reference model could also be chosen as a linear force-free field such as that in \citet{1997Wang} and \citet{2001Wang}. \citet{2003Moon} proposed a uniform shear angle method by assuming that the transverse magnetic field makes an acute angle with the azimuth angle of the potential field transverse component plus an additional shear angle. The magnetic pressure gradient method \citep{1993Cuperman} assumes the magnetic field to be force-free and the magnetic pressure decreases with height. \citet{2004Georgoulis} introduced a structure minimization method by minimizing the electric current density generated by the magnetic field gradients. \citet{2005Georgoulis} further proposed a nonpotential magnetic field calculation method for removing the ambiguity. It employs an iterative method to determine the azimuth by minimizing the discrepancy between the observations and the field computed by the nonpotential model. This method has been improved as described in \citet{2006Metcalf}. In the pseudo-current method \citep{1995Gary}, a multidimensional conjugate gradient method is used to minimize the square of the vertical current density, by which the azimuth is determined. Another iterative method developed in University of Hawaii was described in \citet{1993Canfield}, where the azimuth is initially determined by the acute angle method and the potential or linear force-free field model, and then by the minimization of the square of the vertical current density and the divergence of the magnetic field. The minimum energy method \citep{1994Metcalf,2006Metcalf} employs the simulated annealing method to minimize a function of the magnetic field divergence and the total electric current density, which are derived by a linear force-free field model. This method has been improved by \citet{2009Leka}. Finally, there is an interactive procedure, AZAM, developed in High Altitude Observatory (HAO), to remove the ambiguity by imposing smoothness and matching the magnetic field to expectations of the solar structure \citep{2006Metcalf}.

\citet{2006Metcalf} provided a comprehensive comparison of all the aforementioned ambiguity-removal methods to test their performances. The authors adopted two reference models to get the vector magnetic field, one from an MHD simulation and the other from a theoretical model computed from multi-point sources buried under a plane. They applied each of the method to the reference model and compare the results with it to determine the area, magnetic flux, and strong horizontal field that have been recovered. It shows that those methods minimizing the electric current density and the magnetic field divergence provide the most promising results, such as the nonpotential magnetic field calculation method, the iterative method developed in University of Hawaii, the minimum energy method, and the interactive method developed in High Altitude Observatory. Figure~\ref{fig:proj}a shows a vector magnetic field observed by \textit{SDO}/HMI, whose 180$^\circ$ ambiguity is removed by the minimum energy method. Developing more reliable and effecient methods to remove the 180$^\circ$ ambiguity is still an ongoing topic. Some new algorithms have been discussed in, e.g., \citet{2007Li} and \citet{2009Crouch}.

\subsection{Correction of the Projection Effect} \label{sec:projection}

Due to the spherical nature of the solar surface, vector magnetic field observed on the photosphere is subjected to the projection effect except at the solar disk center. The projection has two effects. On the one hand, it projects the vector magnetic field components into an observer-preferred coordinate system, while the Sun itself prefers a heliographic coordinate system. On the other hand, observations on the image plane distort the geometrical positions of the vector magnetic field on the solar surface. The observer-preferred coordinate system is defined by the image plane and the line of sight, where the $\zeta$ axis is towards the observer, the $\xi$ and $\eta$ axes on the image plane, and the three axes are perpendicular to each other. Additionally, the $\xi$ axis is towards the west and $\eta$ to the north extremity of the earth rotation axis. The heliographic coordinate system is defined by the solar longitudinal direction $x$ (towards the west), the latitudinal direction $y$ (towards the solar north), and the radial direction $z$.

The image plane components ($B_\xi, B_\eta, B_\zeta$) can be transformed to the heliographic components ($B_x, B_y, B_z$) by the following operation \citep{1990Gary}:
\begin{equation}
\left( \begin{array}{c}
B_x(\xi,\eta) \\ B_y(\xi,\eta) \\ B_z(\xi,\eta) \end{array} \right)
=
\begin{array}{c}
 \\
\mathscr{R}(P,B,B_0,L,L_0) \\
\\
\end{array}
\left( \begin{array}{c}
B_\xi(\xi,\eta) \\ B_\eta(\xi,\eta) \\ B_\zeta(\xi,\eta) \end{array} \right) . \label{eqn:r}
\end{equation}
The rotation matrix $\mathscr{R}$ has $3\times 3$ elements and it is a function of $P, B, B_0, L$, and $L_0$, where the solar $P$ angle is the position angle of the north extremity of the solar rotation axis relative to the north extremity of the earth rotation axis, $B$ and $L$ are the latitude and longitude of the vector magnetic field at the image plane coordinate $(\xi, \eta)$, and $B_0$ and $L_0$ are the latitude and longitude of the solar disk center. The rotation matrix can be derived by four successive elementary rotations, namely, rotation around the $\zeta$ axis by angle $P$:
\begin{equation}
\begin{array}{c}
 \\ \mathscr{R}_\zeta(P) \\ \\ \end{array}
=
\left( \begin{array}{ccc}
\cos P  & \sin P & 0 \\
-\sin P & \cos P & 0 \\
0       & 0      & 1 \end{array} \right) , \label{eqn:rz}
\end{equation}
rotation around the $\xi$ axis by $B_0$:
\begin{equation}
\begin{array}{c}
 \\ \mathscr{R}_\xi(B_0) \\  \\ \end{array}
=
\left( \begin{array}{ccc}
1  & 0 & 0 \\
0  & \cos B_0 & \sin B_0 \\
0  & -\sin B_0& \cos B_0  \end{array} \right) , \label{eqn:rx}
\end{equation}
rotation around the $\eta$ axis by $L-L_0$:
\begin{equation}
\begin{array}{c}
 \\ \mathscr{R}_\eta(L - L_0) \\  \\ \end{array}
=
\left( \begin{array}{ccc}
\cos (L-L_0) & 0 & -\sin (L-L_0) \\
0  & 1 & 0 \\
\sin (L-L_0) & 0 & \cos (L-L_0) \end{array} \right) , \label{eqn:ry}
\end{equation}
and rotation around the $\xi$ axis by $-B$, where $\mathscr{R}_\xi(-B)$ is obtained by substituting $B_0$ with $-B$ in Equation~(\ref{eqn:rx}). The rotation matrix for Equation~(\ref{eqn:r}) is then expressed as:
\begin{equation}
\mathscr{R}(P,B,B_0,L,L_0) = \mathscr{R}_\xi(-B) \mathscr{R}_\eta(L - L_0) \mathscr{R}_\xi(B_0) \mathscr{R}_\zeta(P) . \label{eqn:r2}
\end{equation}
The full development of Equation~(\ref{eqn:r2}) can be found in \citet{1990Gary}. Equation~(\ref{eqn:r2}) is easier to understand and debug in a code than the fully developed expression in \citet{1990Gary}.

For the geometrical projection correction, \citet{1990Gary} proposed a linear approximation using a plane tangent to the solar surface at the image center point $(B_c, L_c)$. This approximation omits the curvature of the Sun and assumes a plane as the geometry. Therefore, it is suitable for a Cartesian coordinate system but only applies to a small field of view. The error arising from the plane assumption depends on the size of the field of view. \citet{1990Gary} has estimated such an error quantitatively. The coordinates $(x, y)$ on the de-projected plane map to the coordinates $(\xi, \eta)$ on the image plane with the following linear transformation \citep{1990Gary}:
\begin{equation}
\left( \begin{array}{c}
\xi \\ \eta \end{array} \right)
=
\left( \begin{array}{cc}
c_{11} & c_{12} \\
c_{21} & c_{22} \end{array} \right)
\left( \begin{array}{c}
x \\ y \end{array} \right) . \label{eqn:projection1}
\end{equation}
Both coordinates are referred to the center of the image $(B_c,L_c)$ and the matrix elements are also derived by four successive rotations:
\begin{equation}
\mathscr{R}(P,B_c,B_0,L_c,L_0) = \mathscr{R}_z(-P) \mathscr{R}_x(-B_0) \mathscr{R}_y(L_0-L_c) \mathscr{R}_x(B_c) . \label{eqn:r3}
\end{equation}
Only the upper left four elements in Equation~(\ref{eqn:r3}) are needed to calculate the coefficients in Equation~(\ref{eqn:projection1}). The full development of Equation~(\ref{eqn:r3}) can also be found in \citet{1990Gary}.

\begin{figure}
\begin{center}
\includegraphics[width=1.0\textwidth]{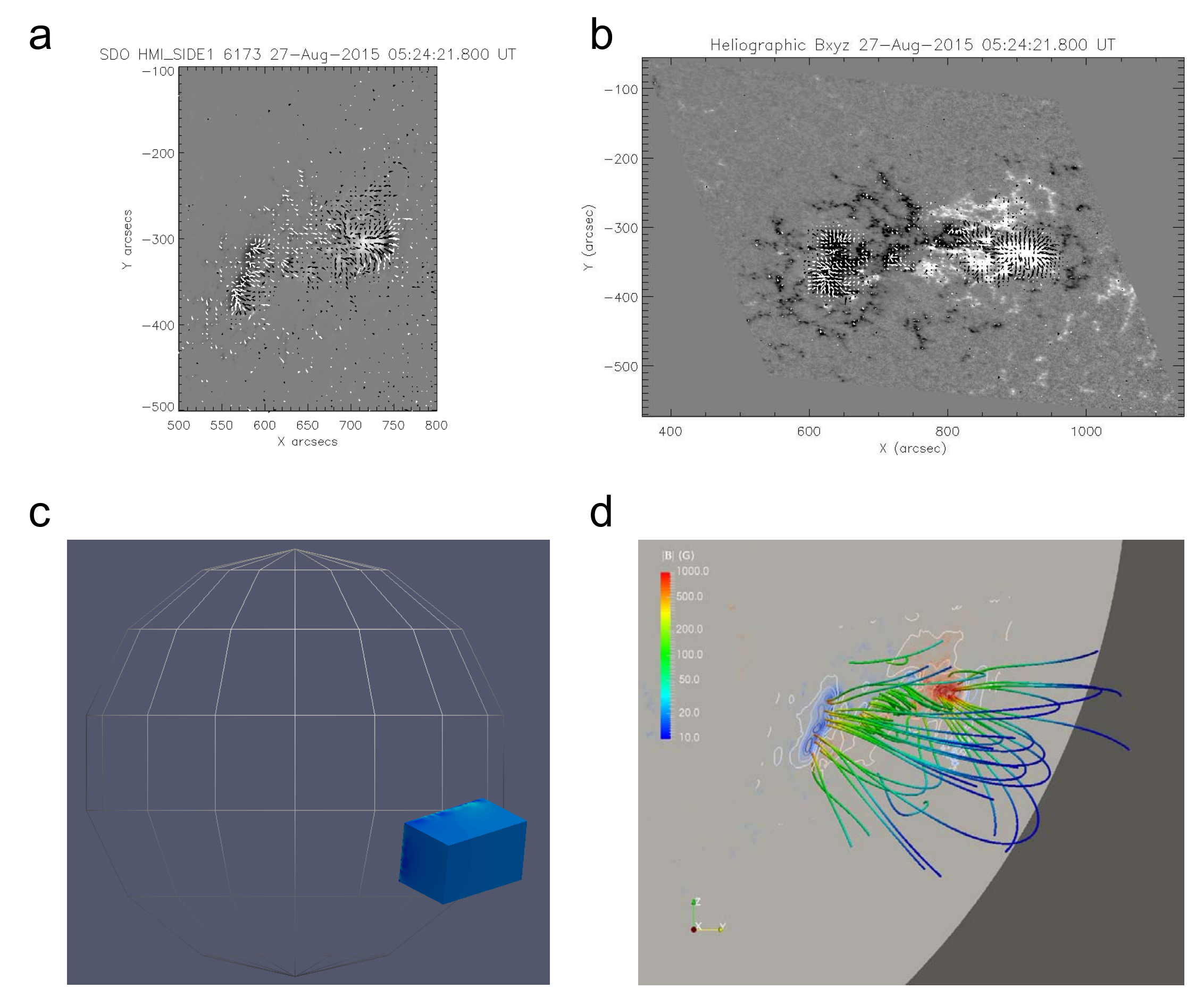}
\caption{A vector magnetic field and nonlinear force-free field model illustrating how the projection effect is corrected. \textbf{a.} Vector magnetic field at 05:24 UT on 2015 August 27 observed by \textit{SDO}/HMI. \textbf{b.} The vector magnetic field components have been transformed to the heliographic coordinate system. Its geometry has been mapped onto a plane tangent to the solar surface at (W54.7$^\circ$, S15.3$^\circ$). \textbf{c.} The computation box of the nonlinear force-free field is rotated back to the place tangent to the solar surface at (W54.7$^\circ$, S15.3$^\circ$). \textbf{d.} The vector magnetic field components are transformed back to the line-of-sight and plane-of-sky coordinate system, where the $x$-axis points to the observer. Solid lines colored with the field strength represent the magnetic field lines. The background image shows the line-of-sight magnetic field, which is the same as that in panel \textbf{a}. White lines show the contours of the local vertical component of the magnetic field, namely, $B_z$ as shown in panel \textbf{b}.
} \label{fig:proj}
\end{center}
\end{figure}

Figure~\ref{fig:proj}b shows an example where the projection effect of the originally observed vector magnetic field in Figure~\ref{fig:proj}a has been corrected. The vector magnetic field components have been transformed to the heliographic coordinate system, and the geometry has been mapped to a plane tangent to the solar surface at the center of the field of view of Figure~\ref{fig:proj}a, which is (W$54.7^\circ$, S$15.3^\circ$). Note that the de-projected vector magnetic field in Figure~\ref{fig:proj}b is equivalent to the result if placing the region of interest at the solar disk center. We could also do the reverse operation of the rotations in Equations~(\ref{eqn:r2}) and (\ref{eqn:r3}) to project the de-projected magnetic field back to the original location. This reverse operation could be done for 3D data as shown in Figure~\ref{fig:proj}c. Figure~\ref{fig:proj}d shows that the inverse operation of the de-projected data quantitatively matches the original observation.

The geometrical projection correction using Equation~(\ref{eqn:projection1}) is limited to a small field of view. When the field of view is large, the plane approximation causes large errors due to the spherical surface of the Sun. This issue has been considered in the HMI data products \citep{2014Hoeksema}, where the vector magnetic field is mapped onto the cylindrical equal area (CEA) coordinate system \citep{2002Calabretta,2006Thompson}. Note that the CEA coordinate system is essentially a spherical coordinate system, and the grids in the latitude are distributed with equal spacing of the sine of the latitude. Therefore, the HMI vector magnetic field data mapped on the CEA coordinate system cannot be used directly as the boundary condition in a problem solved in a Cartesian coordinate system, especially when the field of view is large or the latitude is high.

\subsection{Velocity Derived from Optical Flow Techniques} \label{sec:velocity}
The velocity of magnetized plasma in the photosphere is a critical physical parameter to determine the evolution of the plasma and magnetic field. It can be used as the boundary condition of data-driven MHD simulations and can be used to compute the injection of magnetic energy and helicity. The ling-of-sight velocity can be observed by the Doppler effect, while the vector velocity (in some models, only its transverse components) can be computed by the optical flow techniques, which use a series of magnetograms to infer the velocity with an implied physical evolution model. Traditionally, the local correlation tracking (LCT) method is adopted using only the normal component of magnetograms to derive the photospheric velocity, $\mathbf{u}_\mathrm{LCT}$. This velocity is regarded as the horizontal plasma velocity, $\mathbf{v}_t$, such as in \citet{2001Chae}, \citet{2002Moon}, and \citet{2002Nindos}. However, \citet{2005Schuck} pointed out that the LCT method implies an advection model. It is inconsistent with the magnetic induction equation, which is a continuity equation and governs the evolution of magnetic fields.

\citet{2005Schuck} showed that the LCT method aims to maximize the correlation coefficient of two images in a window. Thus, LCT requires the intensity, $I$, of an image at a given position and time to satisfy the following equation
\begin{equation}
I(\mathbf{x},t_2) \equiv I[\mathbf{x}-\mathbf{u}_0(t_2-t_1),t_1] ,
\end{equation}
which leads to the advection equation
\begin{equation}
\dfrac{\partial I}{\partial t} + \mathbf{u}_0 \cdot \nabla I = 0 ,
\end{equation}
where $I$ is differentiable. While the evolution of the normal component of the vector magnetic field, $B_n$, is governed by the magnetic induction equation. In the photosphere, the resistivity can be omitted such that the plasma is deemed as ideal, and the magnetic induction equation for $B_n$ is written as \citep{2002Kusano,2003Demoulin,2004Welsch,2004Longcope,2006Georgoulis}
\begin{equation}
\renewcommand{\arraystretch}{2.0}
\begin{array}{l}
\dfrac{\partial B_n}{\partial t} + \nabla_t \cdot (B_n
\mathbf{v}_t - v_n \mathbf{B}_t) = 0 ,  \label{eqn:indu3}
\end{array}
\end{equation}
where $n$ denotes the normal direction and $t$ the transverse direction.

Inversion of the velocity from Equation~(\ref{eqn:indu3}) suffers from two ambiguities. On the one hand, \citet{2004Welsch} and \citet{2004Longcope} showed that the flux transport vector can be expressed as two scalar functions, namely, the inductive potential, $\phi$, and the electrostatic potential, $\psi$, as
\begin{equation}
\mathbf{u} B_n = B_n \mathbf{v}_t - v_n \mathbf{B}_t = -(\nabla_t
\phi + \nabla_t \psi \times \hat{\mathbf{n}}) . \label{eqn:ftvel1}
\end{equation}
Only $\phi$ can be determined by Equation~(\ref{eqn:indu3}) while $\psi$ could be arbitrary. Therefore, the velocity can only be determined with additional assumptions, one of which is the assumption of the affine velocity profile in a window \citep{2005Schuck} such that
\begin{equation}
\renewcommand{\arraystretch}{1.2}
\mathbf{u}(\mathbf{x}) = \left( \begin{array}{l} U_0 \\ V_0
\end{array} \right) + \left( \begin{array}{l l} U_x & U_y \\ V_x &
V_y \end{array} \right) \left( \begin{array}{l} x \\ y
\end{array} \right) ,
\end{equation}
with $(U_0, V_0)$ the velocity at the central position, and $U_x$, $U_y$, $V_x$, and $V_y$ the first order spatial derivatives of the velocities. The velocities are derived by a least-square fitting of the model to observations. This method has been implemented as the differential affine velocity estimator (DAVE) in \citet{2006Schuck}.

On the other hand, the plasma velocity along a field line cannot be determined only by the evolution of $B_n$. \citet{2008Schuck} tackled this problem by extending DAVE for vector magnetograms (DAVE4VM), where the velocity is assumed to exhibit a 3D affine velocity profile with
\begin{equation}
\renewcommand{\arraystretch}{1.2}
\mathbf{u}(\mathbf{x}) = \left( \begin{array}{l} U_0 \\ V_0 \\ W_0
\end{array} \right) + \left( \begin{array}{l l} U_x & U_y \\ V_x &
V_y \\ W_x & W_y \end{array} \right) \left( \begin{array}{l} x \\
y
\end{array} \right) .
\end{equation}
The vector velocity is derived by fitting the model to observations of a time sequence of vector magnetic field.

\section{Three-Dimensional Magnetic Field Models} \label{sec:extrapolation}
The magnetic field observation is less accessible in the chromosphere and corona than in the photosphere due to a lower density and a higher temperature. A higher temperature widens the spectral lines that cover the Zeeman splitting. Moreover, the magnetic field is usually weaker at higher altitudes. These problems are alleviated by adopting infrared spectral lines \citep[e.g.,][]{2000Lin,2004Lin,2009Kuckein,2012Kuckein}, because the Zeeman splitting is proportional to the wavelength squared and magnetic field strength. Nevertheless, there are still some difficulties. The line emission in the chromosphere and corona is weaker than that in the photosphere, which lowers the signal-to-noise ratio. Moreover, the optical thin condition in the corona makes the interpretation of the observations more difficult than that in the photosphere. In addition, the plasma in the chromosphere and corona are in a state of non-local thermodynamic equilibrium. The formation of spectral lines there are more difficult to be interpreted than the photospheric spectral lines, where local thermodynamic equilibrium is valid.

Despite the aforementioned difficulties, observations of the magnetic field in the chromosphere have made progress in the past years \citep{2015Lagg,2016DeLaCruz}. The way is to measure the polarization signal of chromospheric spectral lines generated by the Zeeman and Hanle effect. Some typical magnetic sensitive lines in the chromosphere include the He \textsc{i} 10830~\AA \ triplet line and the Ca \textsc{ii} 8542~\AA \ triplet line. Spectropolarimetric observations and line formation theories of the aforementioned spectral lines have been advanced to enable the detection of the chromospheric magnetic field \citep{2000Socas-Navarro,2008Asensio,2009Lagg,2013DeLaCruz}. More explanations and applications of the Zeeman and Hanle effects to detecting the chromospheric and coronal magnetic field can be found in \citet{2013Stenflo} and \citet{2014Schmieder}. Magnetic field in the corona can also be inferred by radio observations \citep[e.g.,][]{2002White,2017Wang}. However, only the strength can be obtained, while the full magnetic vector cannot be derived at present.

Although we have the aforementioned methods, reliable measurements of the full 3D magnetic field is still unavailable. Thus, people resort to theoretical and numerical models to study the solar magnetic field, which include theoretical force-free field models, numerical nonlinear force-free field models, MHS models, and MHD models. Each model has its own advantage and disadvantage, while they may complement each other when combined in a study. For example, theoretical force-free field models have precise solutions, whose magnetic tension and pressure forces balance each other exactly, while they are too simple to describe the complex magnetic structures and dynamical evolution revealed in real observations. Numerical nonlinear force-free field models have the advantage of being related with magnetic field observations directly. However, they still cannot explain the forced structures in the photosphere and lower chromosphere and their dynamical evolution. MHS models partly overcome the disadvantage of the force-free field models, while they still belong to static models. MHD models may provide the best way to study the observed complex magnetic structures and dynamical evolution though they are the most complicated. At present, it is still a difficult task to combine MHD models with other observed parameters like magnetic field, density, and temperature.

\subsection{Theoretical Force-Free Field Models} \label{sec:theory}

The upper chromosphere and lower corona, especially in an active region, are dominated by the magnetic pressure and tension \citep{2001Gary}. The gradient of the gas pressure and the gravity there are much smaller than the magnetic pressure and tension forces. This implies that, when modeling these atmospheric layers, even a very small Lorentz force can break down any static solutions. Therefore, in static models, one always assumes that the Lorentz force equals zero, namely $\mathbf{J} \times \mathbf{B}=0$. The magnetic field without any Lorentz force is defined as a force-free field. The solution requires the electric current density to be parallel to the magnetic field. Since $\mu_0 \mathbf{J} = \nabla \times \mathbf{B}$, we have
\begin{equation}
\nabla \times \mathbf{B} = \alpha \mathbf{B}, \label{eqn:ff}
\end{equation}
where $\alpha$ is a scalar function of the 3D space. Meanwhile, the magnetic field satisfies the solenoidal condition
\begin{equation}
\nabla \cdot \mathbf{B}=0. \label{eqn:df}
\end{equation}
If $\alpha$ is a constant in the whole space, which represents the linear force-free field (where a special case is the potential field when $\alpha=0$), Equations~(\ref{eqn:ff}) and (\ref{eqn:df}) are linear, and they can be solved by a Green's function method or a Fourier transform method in the Cartesian coordinate system \citep{1964Schmidt,1972Nakagawa,1977Chiu,1977Teuber,1978Seehafer}. For a potential field in the spherical coordinate system, the spherical harmonic transformation method is often adopted for solving the Laplace's equation, $\nabla^2 \Phi=0$, for the magnetic field potential $\Phi$, and $\mathbf{B}=-\nabla \Phi$ \citep{1969Schatten,1969Altschuler,1977Altschuler,2003Schrijver}. Other ways of solving the Laplace's equation include finite differences \citep{2011Toth} and a fast solver by combining spectral and finite-difference methods \citep{2012Jiang3}.

For a nonlinear force-free field where $\alpha$ varies in the space, it is difficult to find analytic solutions for Equations~(\ref{eqn:ff}) and (\ref{eqn:df}) with given boundary values, because of the nonlinearity of the problem. Some special force-free field solutions are discussed as follows, and the numerical methods for solving nonlinear force-free field with given boundary conditions are to be discussed in Section~\ref{sec:nlfff}.

A simple but intuitive way to construct potential fields is the magnetic charge method. Some magnetic point sources are placed below a selected surface at $z=0$, and the magnetic field in the space $z>0$ is assumed to be potential. Thus, the magnetic field can be derived similarly as the electric field derived by the Coulomb's law:
\begin{equation}
\mathbf{B} = \sum_i \frac{m_i \mathbf{r}_i}{r_i^3}, \label{eqn:mc}
\end{equation}
where $i$ is the number of the magnetic charges, $m_i$ is the strength of the magnetic charge, $\mathbf{r}_i$ is the position vector pointing from the magnetic charge $i$ to the place of magnetic field, and $r_i$ is the scalar distance between them. Potential fields computed by the magnetic charge method have been adopted to study various aspects of magnetic structures and solar activities \citep{1986Seehafer,1988Gorbachev,1992Demoulin2,2016Pontin}. Besides the potential field, the magnetic charge method can also be used to construct linear force-free field \citep{1992Demoulin}.

Using an axis symmetric assumption, \citet{1990Low} provided a class of nonlinear force-free field solution in the spherical coordinate system. An axis symmetric and solenoidal magnetic field $\mathbf{B}$ can be written as
\begin{equation}
\mathbf{B} = \frac{1}{r \sin \theta}\left (\frac{1}{r}\frac{\partial A}{\partial \theta} \mathbf{e}_r, -\frac{\partial A}{\partial r} \mathbf{e}_\theta, Q \mathbf{e}_\phi \right ), \label{eqn:ll}
\end{equation}
where $A$ and $Q$ are functions of only $r$ and $\theta$ because of the symmetry. Substituting Equation~(\ref{eqn:ll}) into Equation~(\ref{eqn:ff}), one could find that $Q$ is a function of $A$, and $A$ satisfies the Grad-Shafranov equation, which is a two-dimensional (2D) nonlinear partial differential equation \citep{1958Grad}. \citet{1990Low} further found a class of separable solutions of $A$. The problem is finally reduced to a nonlinear ordinary differential equation of a scalar function, which is solvable with a numerical method. Thus, the Low and Lou solution is often called semi-analytic. It has been widely adopted as a reference model for testing the numerical nonlinear force-free field algorithms. Figure~\ref{fig:mf}a shows one of such tests.

Another class of semi-analytic nonlinear force-free field solution is the flux rope model proposed by \citet{1999Titov}. The Titov--D\'emoulin model include a nonlinear force-free field assumption and concentration of the electric current into a partial torus above a selected surface. \citet{1999Titov} argued that the general magnetic topology is determined by magnetic field lines, which are a double integration of the electric current distribution. This effect smoothes out the small-scale and weak current distribution and only leaves the large-scale and strong one that is essential to the problem. In this way, a single current channel in a torus shape could characterize the main feature of an active region with twisted magnetic field lines. The hoop force of the current torus is balanced by the Lorentz force caused by the interaction between the current and the potential field, which is generated by two imaginary magnetic charges and a line current that are buried under the selected surface. An example of the Titov--D\'emoulin model is provided in Figure~\ref{fig:mf}b.

Some other analytic solutions of force-free fields can be found in, e.g., \citet{1960Gold}, \citet{2011Titov}, \citet{2014Kleman}, and \citet{2015Kleman}. All these models provide various insights into the magnetic field topology and stabilities.

\subsection{Numerical Nonlinear Force-Free Field Models} \label{sec:nlfff}

In order to extract the full information from vector magnetic field observations, one has to resort to numerical methods that aim at reconstructing nonlinear force-free field models from boundary conditions and proper pseudo initial conditions. The latter serve as the initial state for an iteration process, and the final converged results are usually independent of them. Compared to a potential field model, a nonlinear force-free field involves free magnetic energy and electric current, which could power solar eruptions. When compared with a linear force-free field model, a nonlinear force-free field is closer to coronal observations when there are both large scale coronal loops and low-lying filaments or prominences. Thus, it is closer to the realistic magnetic configuration. As a result, a nonlinear force-free field provides a better initial condition for further MHD simulations than the potential and linear force-free field models. The nonlinear force-free field model has also been adopted to reconstruct magnetic flux ropes in the solar atmosphere \citep{2009Canou,2010Canou,2009Su,2010Guo,2010Guo2,2010Cheng,2010Jing,2014Jiang2}.

There are various numerical algorithms to compute nonlinear force-free field models, such as the Grad--Rubin, vertical integration, MHD relaxation, optimization, and boundary integral equation methods. Here, we provide a brief introduction of these methods. Readers are referred to \citet{2012Wiegelmann} for more details.

The Grad-Rubin method was first proposed by \citet{1958Grad} for computing nonlinear force-free field in fusion plasma. The idea is to compute the torsional parameter $\alpha$ and magnetic field $\mathbf{B}$ iteratively. Given an initial condition $\mathbf{B}^{(0)}$ by the potential magnetic field, $\alpha$ at each iteration step $n$ is computed by a hyperbolic equation
\begin{equation}
\mathbf{B}^{(n)} \cdot \nabla \alpha^{(n)} = 0. \label{eqn:grad1}
\end{equation}
The boundary condition of $\alpha$ is provided at either the positive or the negative polarity, which can be derived from the vector magnetic field. Then, the magnetic field at the next step $n+1$ is computed by solving an elliptic equation
\begin{equation}
\nabla \times \mathbf{B}^{(n+1)} = \alpha^{(n)} \mathbf{B}^{(n)}. \label{eqn:grad2}
\end{equation}
The newly computed magnetic field $\mathbf{B}^{(n+1)}$ is also constrained by the solenoidal condition and the boundary condition for the normal component. The Grad--Rubin method was first applied to compute coronal magnetic fields by \citet{1981Sakurai}, and it has been further developed and applied to magnetic field reconstruction by \citet{1997Amari,2006Amari,2014Amari}, \citet{2004Wheatland,2006Wheatland}, \citet{2006Inhester}, and \citet{2013Gilchrist,2014Gilchrist}. A modified Grad--Rubin method has also been proposed by \citet{2012Malanushenko,2014Malanushenko} using information of the torsional parameter $\alpha$ in the volume, which is derived from the coronal loop geometry. The solenoidal condition of the magnetic field is guaranteed by different strategies in different implementations. For example, \citet{1997Amari,2006Amari,2014Amari} used the vector potential to represent the magnetic field, which could ensure $\nabla \cdot \mathbf{B} = 0$ with the accuracy of round-off errors.

The vertical integration method is straightforward and relatively easy to implement \citep{1974Nakagawa}. It first computes $\alpha$ from a vector magnetic field on the bottom using the $z$ component of the force-free Equation~(\ref{eqn:ff}) as
\begin{equation}
\alpha=\frac{1}{B_z} \left(\frac{\partial B_y}{\partial
x} - \frac{\partial B_x}{\partial y} \right) .
\label{eqn:vert1}
\end{equation}
Then, the vector magnetic field and $\alpha$ higher up are derived by integrating the following equations over $z$
\begin{equation}
\renewcommand{\arraystretch}{2.0}
\begin{array}{c}
\dfrac{\partial B_x}{\partial z} =  \alpha B_y + \dfrac{\partial
B_z}{\partial x}, \\
\dfrac{\partial B_y}{\partial z} = -\alpha B_x + \dfrac{\partial
B_z}{\partial y}, \\
\dfrac{\partial B_z}{\partial z} = -\dfrac{\partial B_x}{\partial
x}-\dfrac{\partial B_y}{\partial y}, \\
\dfrac{\partial \alpha}{\partial z} = -\dfrac{1}{B_z} \left(B_x
\dfrac{\partial \alpha}{\partial x} + B_y \dfrac{\partial
\alpha}{\partial y} \right) . \label{eqn:vert2}
\end{array}
\end{equation}
Unfortunately, Equation~(\ref{eqn:vert2}) is ill-posed in the sense that numerical errors grow fast with height, which has been discussed in \citet{1990Low}, \citet{1990Wu}, and \citet{1992Demoulin3}. Some methods have been proposed to regularize the problem by smoothing the physical variables \citep{1990Cuperman,1992Demoulin3,2006Song}.

The MHD relaxation method solves the MHD momentum equation and magnetic induction equation to build an equilibrium state. The momentum equation includes a dissipative term $\mathbf{D}(\mathbf{v})$, which is a function of the velocity $\mathbf{v}$, and the magnetic induction equation considers the resistive term:
\begin{equation}
\rho(\frac{\partial}{\partial t}+\mathbf{v}\cdot\nabla)\mathbf{v}=
\mathbf{J}\times\mathbf{B}-\nabla p+\rho\mathbf{g}+\mathbf{D(v)}
,\label{eqn:mome}
\end{equation}
\begin{equation}
{\partial {\bf B}\over \partial t} = \nabla \times ({\bf v} \times
{\bf B}) - \nabla \times (\eta \mathbf{J}),
\label{eqn:indu1}
\end{equation}
where $\rho$ is the density, $\mathbf{J} = \nabla \times \mathbf{B}/\mu_0$  is the electric current density, $\mu_0$ is the vacuum permeability, $p$ is the gas pressure, $\mathbf{g}$ is the gravitational acceleration vector, and $\eta$ is the resistivity. Considering that $\eta$ is uniform in the space, Equation~(\ref{eqn:indu1}) can be recast as
\begin{equation}
{\partial {\bf B}\over \partial t} = \nabla \times ({\bf v} \times
{\bf B}) + \eta_\mathrm{m} \nabla^2 \mathbf{B},
\label{eqn:indu2}
\end{equation}
where $\eta_\mathrm{m} = 1/(\sigma \mu_0)$ is the magnetic diffusivity, and $\sigma = 1/\eta$ is the conductivity. The dissipative term $\mathbf{D(v)}$ can be either in a friction form such that $\mathbf{D(v)}=-\nu\mathbf{v}$ \citep{1986Yang,1996Roumeliotis,2005Valori,2007Valori,2010Valori} or in a viscosity form such that $\mathbf{D(v)} = \nabla \cdot (\nu \rho \nabla \mathbf{v})$ \citep{1994Mikic,1994McClymont,1997McClymont,1996Amari,1997Amari}. The free parameter $\nu$ is used to control the dissipative speed.

\begin{figure}
\begin{center}
\includegraphics[width=1.0\textwidth]{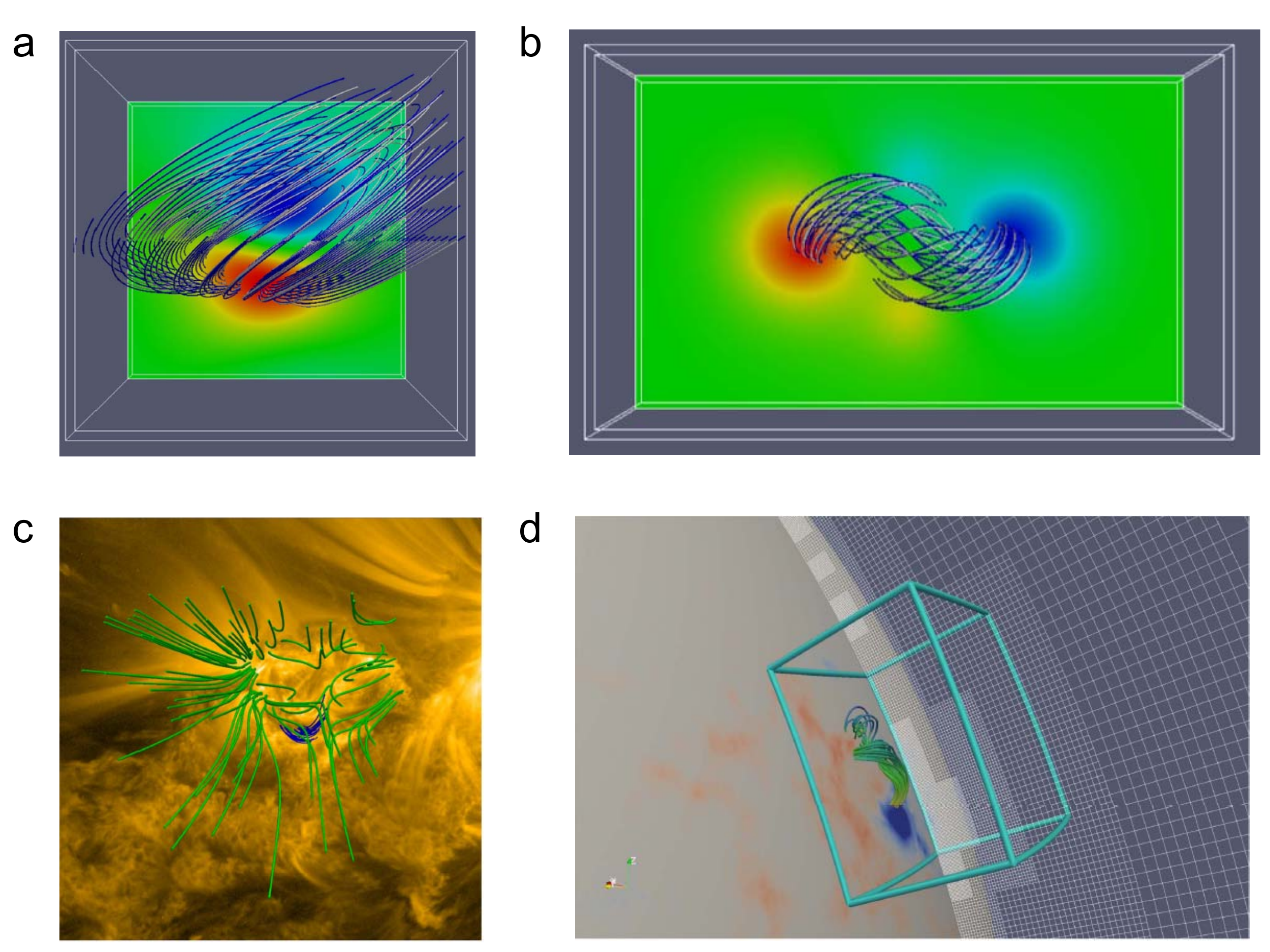}
\caption{Magnetic field models computed by the magneto-frictional method. \textbf{a.} Blue lines show the semi-analytic solution of \citet{1990Low}. White lines are those computed by the magneto-frictional method with all the six boundaries being provided by the original solution. The figure is from \citet{2016Guo1} and reproduced by permission of the AAS. \textbf{b.} Blue lines show the semi-analytic solution of \citet{1999Titov}. White lines are those computed by the magneto-frictional method with all the six boundaries being provided by the original solution. The figure is from \citet{2016Guo1} and reproduced by permission of the AAS. \textbf{c.} Green and blue lines show the nonlinear force-free field model computed by the magneto-frictional method in a Cartesian coordinate system with a uniform grid. It uses a vector magnetic field observed by \textit{SDO}/HMI as the boundary condition. The background image is a 171~\AA \ image by \textit{SDO}/AIA. The figure is from \citet{2016Guo2} and reproduced by permission of the AAS. \textbf{d.} Colored solid lines show a twisted magnetic flux rope modeled by the magneto-frictional method in the spherical coordinate system with an adaptive mesh refinement. The grids show the structure of the adaptive mesh refinement on a slice. The background image represents the radial magnetic field. The figure is adapted from \citet{2016Guo2} and reproduced by permission of the AAS.
} \label{fig:mf}
\end{center}
\end{figure}

In its full form, the MHD relaxation method eventually reaches a state of magneto-hydrostatic equilibrium \citep{1981Chodura,2013Zhu,2016Zhu}. When a nonlinear force-free field model is considered, the inertia, pressure gradient, and gravity forces are omitted in Equation~(\ref{eqn:mome}). If the friction form is adopted, the momentum equation is further reduced to
\begin{equation}
\mathbf{v}=\frac{1}{\nu}\mathbf{J}\times\mathbf{B}
.\label{eqn:mfv1}
\end{equation}
A nonlinear force-free field model could be reconstructed with Equations~(\ref{eqn:indu2}) and (\ref{eqn:mfv1}), which is called as the magneto-frictional method. The initial condition could be a potential magnetic field, and the boundary condition an observed vector magnetic field. Some other initial and boundary conditions are also possible, such as those adopted in the magnetic flux rope insertion method \citep{2004VanBallegooijen,2009Su,2015Savcheva,2016Savcheva}. The flux rope insertion method only uses the normal component of the vector magnetic field on the bottom boundary, and the initial condition is provided by a potential field combined with an artificially inserted magnetic flux rope. Modern MHD codes are recently introduced to nonlinear force-free field reconstructions based on the MHD relaxation approach. For example, \citet{2011Jiang,2012Jiang2} and \citet{2012Jiang1} introduced a CESE-MHD-NLFFF code with the conservation element/solution element scheme \citep{2010Jiang}. The MHD relaxation method has also been developed and applied to a series of researches by \citet{2011Inoue,2012Inoue,2014Inoue}. \citet{2016Guo1} has implemented a new magneto-frictional algorithm in the Message Passing Interface Adaptive Mesh Refinement Versatile Advection Code\footnote{\url{https://gitlab.com/mpi-amrvac}} \citep[MPI-AMRVAC;][]{2003Keppens,2012Keppens,2014Porth} and tested it with analytic solutions. This method is parallelized with MPI and could be applied to both Cartesian and spherical coordinates, with either uniform or adaptive mesh refinement (AMR) grids (Figures~\ref{fig:mf}a, \ref{fig:mf}b, \ref{fig:mf}c, and \ref{fig:mf}d). The solenoidal condition of the magnetic field is guaranteed by including a diffusive term in the induction equation. \citet{2016Guo2} has applied the magneto-frictional method implemented in MPI-AMRVAC to the vector magnetic field observed by \textit{SDO}/HMI in Cartesian coordinates with uniform or AMR grids and in spherical coordinates with AMR grids (Figures~\ref{fig:mf}c and \ref{fig:mf}d).

The optimization method constructs a nonlinear force-free field model by minimizing an objective functional
\begin{equation}
L =  \int\limits_{V} \bigl[B^{-2} \mid ( \nabla \times \mathbf{B} )
\times \mathbf{B}\mid^2 +  \mid \nabla \cdot \mathbf{B} \mid^2
\bigr]~\mathrm{d}V ,\label{eqn:optim}
\end{equation}
where $V$ is the computational volume. When $L$ is minimized to a small value by an iteration process, the force-free and solenoidal conditions are assumed to be fulfilled simultaneously. This method was proposed by \citet{2000Wheatland} and further developed by \citet{2004Wiegelmann} and \citet{2007Wiegelmann} in both Cartesian and spherical coordinate systems. \citet{2009Tadesse,2011Tadesse} further tested and applied the optimization method in spherical geometry to reconstruct the global coronal field. Jim McTiernan developed another version of this method using IDL\footnote{\url{http://sprg.ssl.berkeley.edu/~jimm/fff/optimization_fff.html}} and FORTRAN languages, in both the Cartesian and spherical coordinate systems. Tests and applications can be found in \citet{2006Schrijver} and \citet{2008Metcalf} for the Cartesian version, and in \citet{2012Guo} for the spherical version. Additionally, \citet{2006Inhester} proposed a finite-element scheme for the optimization method, while all the other codes use finite-difference schemes.

The boundary integration method was proposed by \citet{1997Yan,2000Yan}. Using the Green's theorem \citep{1962Courant}, the magnetic field is expressed as a boundary integration as
\begin{equation}
c \mathbf{B} = \int\limits_{S} \left(\mathbf{Y} \dfrac{\partial
\mathbf{B}}{\partial n} - \dfrac{\partial \mathbf{Y}}{\partial n}
\mathbf{B} \right) \mathrm{d} S, \label{eqn:boun1}
\end{equation}
where $c=1/2$ on the boundary $S$ and $c=1$ in the volume above $S$, and $Y$ is a reference function of a diagonal matrix that can be determined by a volume integration. This method has been improved and applied to observations in \citet{2001Yan}, \citet{2006Yan}, and \citet{2006He,2008He}. And recently, it has also been implemented with the acceleration of the graphics processor unit \citep[GPU;][]{2013Wang}.

\subsubsection{Preprocessing of the Vector Magnetic Field for Nonlinear Force-Free Field Extrapolation} \label{sec:prepro}

The vector magnetic field observed on the photosphere is not force-free, since the plasma $\beta$ is close to one there. Thus, inconsistence arises when the forced photospheric field is used as boundary conditions of force-free models. For a force-free field, its boundary value must satisfy the magnetic force-free and torque-free formulae as derived by \citet{1969Molodenskii} and \citet{1989Aly}. Therefore, \citet{2006Wiegelmann} proposed a method to remove the magnetic force and torque using an optimization method to minimize a functional $L$ that represents the sum of the magnetic force, torque, deviation from observations, and the smoothness of the magnetic field. To apply the preprocessing method proposed by \citet{2006Wiegelmann}, some prerequisites need to be satisfied, namely, the vector magnetic field on the bottom boundary should be isolated and close to the disk center. Since the force-free and torque-free conditions are asked to be satisfied on the whole boundary of the computation box, while observations are only available on the bottom, the isolated condition is employed that can allow the force and torque on the lateral and top boundaries to be neglected. An isolated magnetic field means that most of the magnetic flux is concentrated in the field of view and the magnetic flux is balanced. The second prerequisite requires the line-of-sight component to be close to the vertical component, as proposed by \citet{2006Wiegelmann}. This requirement could be loosened by allowing the vertical component to change in a larger range of the uncertainties.

\citet{2007Fuhrmann} developed another method for preprocessing the vector magnetic field based on the principle proposed in \citet{2006Wiegelmann}. This method is different from that of \citet{2006Wiegelmann} in the following three aspects. First, the deviation of the preprocessed magnetic field from the original observations is not included in the functional $L$. Second, the magnetic field is smoothed by a windowed median averaging rather than a 2D Laplacian that is used in \citet{2006Wiegelmann}. Third, a simulated annealing method is adopted for a better convergence to search for the global minimum of the functional $L$ rather than the Newton--Raphson method used in \citet{2006Wiegelmann}.

\citet{2014Jiang} also developed a preprocessing method for removing the magnetic force and torque of the photospheric vector magnetic field. The idea is to split the observed vector magnetic field into a potential part determined only by the vertical component and a non-potential part. They argued that the potential magnetic field at a height of about 400~km, which is approximately the length of a pixel in the HMI data, can be regarded as the preprocessed potential field, while the non-potential part is preprocessed with the same method of \citet{2006Wiegelmann}. The potential field is used to guide the preprocessing of the non-potential part, which is required to possess the same level of force-freeness and smoothness as that of the potential field at the height of 400~km.

All the aforementioned preprocessing methods are developed in the Cartesian coordinate system. A preprocessing method in the spherical geometry has been developed \citep{2007Wiegelmann,2009Tadesse,2011Tadesse}, which is also implemented and tested by \citet{2012Guo} using the formulae in \citet{2009Tadesse}. A series of tests show that preprocessing can improve the nonlinear force-free field extrapolation by decreasing the magnetic divergence and Lorentz force \citep{2006Wiegelmann,2007Fuhrmann,2008Metcalf,2011Fuhrmann,2012Guo,2013Jiang1}. Note that the preprocessing has to modify the photospheric data to satisfy the force-free condition and the resulted magnetic field is assumed to correspond to the chromospheric field. The validity of this argument is yet to be checked by direct observations of the chromospheric magnetic field or more sophisticated models.

\subsection{Magnetohydrostatic Models} \label{sec:mhs}

Different from the aforementioned nonlinear force-free field models, a series of non-force-free models have also been developed. The models are magnetohydrostatic (MHS) in essence, which include the physical effects of pressure gradient and gravity but omit the plasma inertia:
\begin{equation}
\mathbf{J}\times\mathbf{B}-\nabla p+\rho\mathbf{g} = 0. \label{eqn:mhs}
\end{equation}
There are two different ways to construct an MHS model governed by Equation~(\ref{eqn:mhs}), one being analytic and the other numerical. To use the analytic method, additional assumptions are needed to simplify the governing equations of the MHS model to linear problems. For example, assuming electric currents to be perpendicular to gravity everywhere in the computational volume, \citet{1985Low} and \citet{1986Bogdan} found a class of analytic solutions for Equation~(\ref{eqn:mhs}). With the same assumption as \citet{1985Low}, \citet{1993Zhao,1994Zhao} showed that the solutions of the MHS model can be expressed as the summation of involving spherical harmonics. \citet{1995Neukirch} proposed a new mathematical procedure to calculate the analytic solutions of the MHS equations. Different from \citet{1985Low} and \citet{1986Bogdan}, these solutions allow additional field-aligned electric currents. The solutions of \citet{1995Neukirch} have been applied to modeling a polar crown soft X-ray arcade \citep{2000Zhao} and global coronal structures \citep{2008Ruan}. Using the principle of minimum dissipation rate, \citet{2008Hu1} and \citet{2008Hu2} showed that a general non-force-free field can be expressed as the summation of two linear force-free fields and one potential field. The two free parameters of constant $\alpha$ are determined by comparing the model and the observed transverse magnetic field on the photosphere.

Without any prerequisite on the solutions of the MHS equation, the problem is fully nonlinear and can only be solved by a numerical method. \citet{2003Wiegelmann} proposed to use the optimization method, which is similar to that for the nonlinear force-free field model but including the pressure gradient force and gravity, to construct MHS models. \citet{2007Wiegelmann2} further extended this optimization method for MHS models from Cartesian coordinates to spherical geometry. Another numerical method to construct MHS models is the MHD relaxation method, which is also similar to that for the nonlinear force-free field model but including the pressure gradient and gravity. \citet{1981Chodura} applied the MHD relaxation method to reconstruct MHS models. This method has been further implemented and applied to observations by \citet{2013Zhu,2016Zhu}.

\subsection{Magnetohydrodynamic Models} \label{sec:mhd}

The nonlinear force-free field and MHS models mentioned above are all static models. To study the dynamics of the magnetic field and its interaction with the plasma, we need MHD numerical simulations. Depending on different criteria, MHD simulations are divided into different categories. First of all, MHD simulations can be divided into zero-$\beta$, isothermal, ideal, resistive, and full MHD models with the order of increasing physical details included.

The zero-$\beta$ MHD model omits gravity and gas pressure in the momentum conservation equation, and omits the energy conservation equation. The magnetic diffusion term in the magnetic induction equation could either be neglected \citep{2003Torok,2013Kliem} or not \citep{2010Aulanier}. An isothermal model considers gravity and gas pressure in the momentum conservation equation but the temperature is kept unchanged; thus the energy conservation equation is neglected \citep{2014Xia}. In the ideal MHD model, all the aforementioned physical terms and equations are solved except for the omission of the magnetic diffusion term in the magnetic induction equation. Meanwhile, the thermal conduction, radiative loss, viscous dissipation, and Joule heating are not considered in the energy conservation equation \citep{2009Fan}. A resistive MHD model includes magnetic resistivity in the magnetic induction equation and Joule heating in the energy conservation equation \citep{2013Leake,2014Leake}. A full MHD model tends to include all the physical effects, especially the thermal conduction and radiative loss \citep{2014Xia2}. But in practice, magnetic field diffusion in the induction equation, viscous dissipation and Joule heating might be neglected to save computational time.

Secondly, different MHD models have different physical domains of simulation. They can be restricted only to the solar corona \citep{2003Torok,2012Roussev,2013Kliem}, or a more complete domain that includes also the photosphere, chromosphere, and the convection zone below the solar surface \citep{2001Fan,2009Fan,2001Magara,2003Magara,2004Archontis,2008Archontis}, or an even larger domain extending from the solar atmosphere to the interplanetary space \citep[e.g.,][]{2014Shen,2015Feng}.

\begin{figure}
\begin{center}
\includegraphics[width=1.0\textwidth]{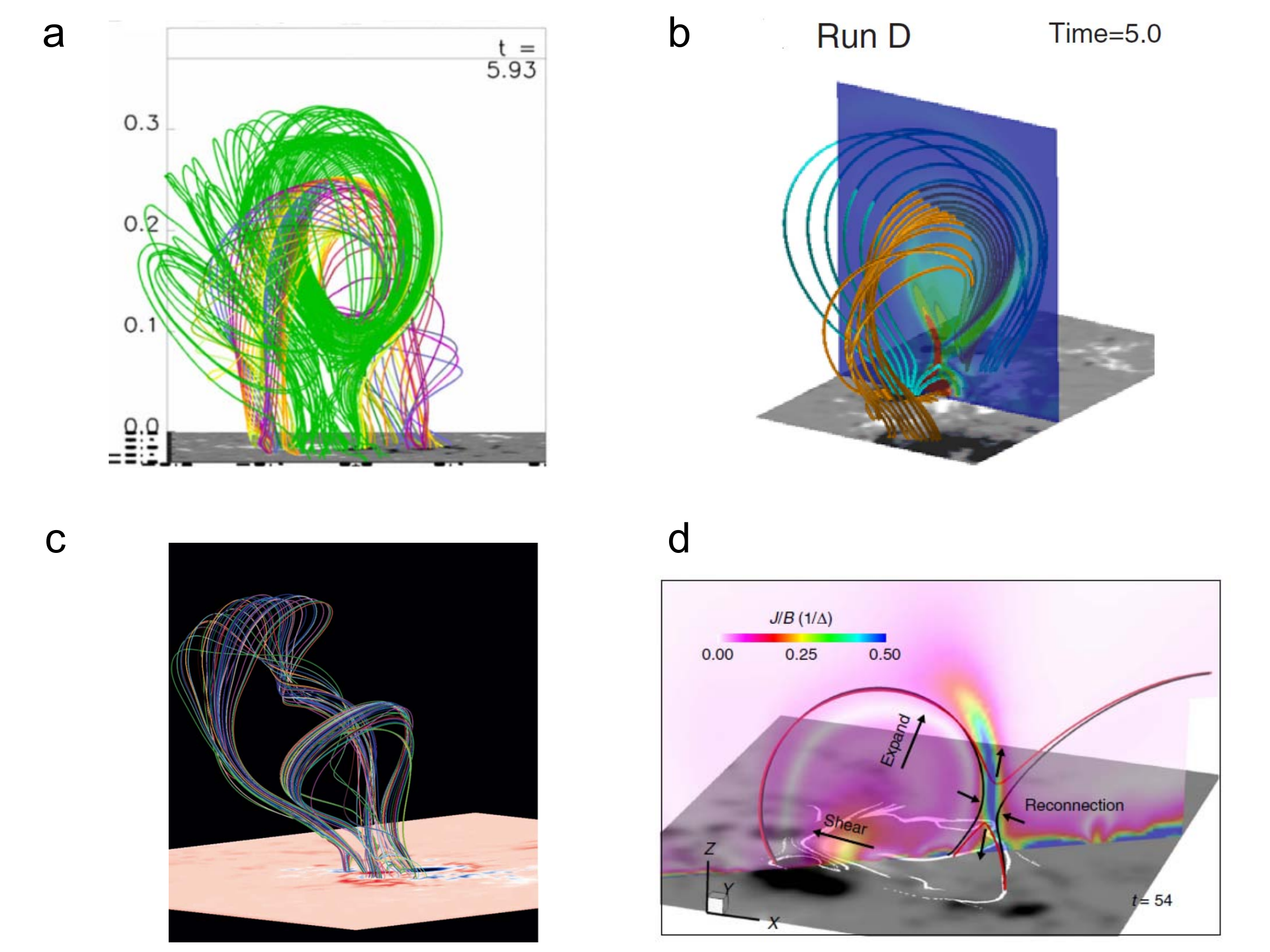}
\caption{Some selected data-constrained and data-driven MHD models. \textbf{a.} A snapshot of the zero-$\beta$ MHD numerical simulation from \citet{2013Kliem} and reproduced by permission of the AAS. Green lines show the overlying magnetic field lines and other colored lines show a twisted magnetic flux rope. \textbf{b.} Another zero-$\beta$ HMD model from \citet{2014Inoue} and reproduced by permission of the AAS. Solid lines with different colors show the magnetic flux rope with different twist numbers. the vertical slice shows the distribution of the vertical velocity and the horizontal slice shows the distribution of vertical magnetic field. \textbf{c.} An erupting twisted magnetic flux rope illustrated by the colored lines from \citet{2014Amari} and reprinted by permission from the Nature Publishing Group. \textbf{d.} A jet-like structure showing the magnetic reconnection induced by the data-driven MHD simulation. Black/red lines indicate magnetic field lines before/after the reconnection. The figure is adapted from \citet{2016Jiang}, where more explanation of the figure is provided.
} \label{fig:mhd}
\end{center}
\end{figure}

Finally, in terms of the initial and boundary conditions adopted, MHD simulations are either purely theoretical \citep{2000Amari,2012Zuccarello} or data-driven/data-constrained \citep{2013Jiang,2016Jiang,2013Kliem,2014Amari,2014Inoue,2015Inoue}. A recent comprehensive review has been made by \citet{2016Inoue} on this topic. Here, we define a data-driven model as the one in which both the initial and boundary conditions are provided by observations. Additionally, the boundary condition should be time-varying in correspondence with the data stream from observations. And a data-constrained model is defined as the one in which only the initial condition is provided by observations. Figure~\ref{fig:mhd}a shows a data-constrained model \citep{2013Kliem}, where the initial condition is provided by the nonlinear force-free field modeled with the flux rope insertion method. The velocity at the bottom boundary is kept at zero all the time. Another data-constrained model is shown in Figure~\ref{fig:mhd}b \citep{2014Inoue}. Two data-driven models are presented in Figures~\ref{fig:mhd}c \citep{2014Amari} and \ref{fig:mhd}d \citep{2016Jiang}, where both the initial conditions are provided with a nonlinear force-free field model, while the boundary conditions by information from observations extracted with different strategies. We note that the simulation in \citet{2014Amari} is not purely data-driven, because their boundary condition for the velocity field is artificially specified with converging flows that mimic the flux cancelation to build up the flux rope until its eruption.

MHD simulations have been applied to a vast range of topics in the solar and space physics. Here, we only focus on a few of the topics closely related to the origin and structure of solar eruptions. As a key ingredient of solar eruptions, magnetic flux ropes are present in most MHD simulations related to solar active region formation and solar eruptive activities \citep[e.g.,][]{2001Fan,2001Magara,2012Roussev,2016Jin}. The driving mechanisms for the eruption of a magnetic flux rope have been studied by some MHD simulations. For example, \citet{2004Torok} found that helical kink instability could drive the initial eruption of a highly twisted flux rope. While there is no evidence showing that the helical kink instability itself could drive the full eruption, \citet{2005Torok} found that the decrease of the overlying magnetic field with height should be fast enough to enable the full eruption of a magnetic flux rope.

One possible formation process of magnetic flux ropes in an active region is through magnetic flux emergence from the convection zone. A series of 3D ideal MHD simulations have been applied to study the process of magnetic flux emergence \citep[e.g.,][]{2004Manchester,2006Murray,2007Galsgaard,2008Magara,2009Archontis,2009Fan}. It is thought that the magnetic buoyancy instability makes the magnetic fluxes break through the photosphere into higher layers. Full MHD simulations including radiative transfer have also been applied to study the self-consistent magneto-convection process in the upper convection zone and the photosphere \citep[e.g.,][]{2000Stein,2008Cheung,2008Martinez,2009Rempel,2017Chen}. Comprehensive reviews on the observations and MHD simulations of magnetic flux emergence can be found in \citet{2009Fan2} and \citet{2009Nordlund}.

In a series of zero-$\beta$ MHD simulations, \citet{2012Aulanier,2013Aulanier} and \citet{2013Janvier} have extended the standard flare model to 3D. \citet{2012Aulanier} found both direct and return electric currents in sunspots and faculae. Flare ribbons usually appear as a $J$-shaped structure as shown by many observations. In the 3D model, the straight part of the $J$-shaped ribbon corresponds to the footpoints of the reconnected field lines, which are formed in the vertical current sheet stretched by the erupting magnetic flux rope. Only direct electric currents appear in this part. While the curved part of the $J$-shaped ribbon corresponds to the periphery of the legs of the erupting flux rope \citep[see also][]{2016Cheng}. Both direct and return currents are present there. The strong-to-weak shear transition of the flare loops are explained by two effects: one is the transfer of differential magnetic shear in the pre-eruptive configuration to the post-eruptive one, and the other is the vertical straightening of the inner legs of the erupting magnetic flux rope. Combining the zero-$\beta$ MHD model and historical records of solar active regions, \citet{2013Aulanier} estimated the largest possible flare on the Sun to be $\sim 6 \times 10^{33}$ erg. \citet{2013Janvier} analyzed the slipping velocity, $v_s$, and the norm of the magnetic field line mapping, $N$, in this MHD simulation. They found that $v_s$ and $N$ are linearly correlated with each other. A comprehensive review on 3D models of solar flares can be found in \citet{2015Janvier}.

The formation of solar filaments/prominenes has been simulated with full 3D MHD models in recent years \citep{2014Xia2,2016Xia1,2016Xia2,2015Keppens}. These models focus on a particular physical mechanism for the prominence formation, namely, the chromospheric evaporation and coronal condensation. They are caused by an impulsive heating in the chromosphere and a runaway radiative loss in the corona. Therefore, the thermal conduction and radiative loss in the energy equation are essential for such a model. \citet{2014Xia2} first showed the full 3D MHD simulation of prominence formation in which hot plasma could condense to cold material in the corona. \citet{2016Xia1} studied the plasma circulation between the chromosphere and corona. They found that when plasmas are heated in the chromosphere, they could be evaporated into the corona. Due to the runaway radiative loss in the corona, hot plasmas are condensed and cooled down to form the cold prominence. The dense prominence plasmas in the corona usually move downward, drag magnetic field with them, and finally fall back to the chromosphere. \citet{2015Keppens} and \citet{2016Xia2} further studied the dynamics of the prominence and showed how the magneto-convective motions and Rayleigh-Taylor instability generate the falling fingers and uprising bubbles as observed by \citet{2008Berger} and \citet{2010Berger}.

\section{Magnetic Energy Computation} \label{sec:energy}
In the solar atmosphere, the energy contained in the magnetic field is much larger than that in other forms, such as the kinematic, thermal, and gravitational potential energy. Therefore, the magnetic energy is thought to be the major reservoir for powering solar flares and CMEs. The total magnetic energy in a volume, $V$, is expressed as
\begin{equation}
E = \displaystyle\int\limits_{V} \frac{B^2}{2\mu_0}
\mathrm{d} V . \label{eqn:energy}
\end{equation}
Not all the magnetic energy can be released in the solar atmosphere. Considering that the evolution of the photosphere is much slower than the dynamical eruptions in the corona, only the free energy, $E_\mathrm{f}$, higher than the potential field, $E_\mathrm{p}$, can be released:
\begin{equation}
E_\mathrm{f} = E - E_\mathrm{p} , \label{eqn:Ef}
\end{equation}
where $E_\mathrm{p}$ is determined by the same normal field of $\mathbf{B}$ on the boundary of the volume, $V$. Additionally, the free energy is only an upper limit for the available energy to be released, since the final state of the magnetic field after flares and/or CMEs is also constrained by the conservation of magnetic helicity \citep{1986Taylor}.

Equation~(\ref{eqn:Ef}) is valid with another requirement that the magnetic field should be solenoidal \citep{2013Valori}. This issue matters when we apply Equation~(\ref{eqn:Ef}) to magnetic field derived by numerical force-free, MHS, and MHD models, where the solenoidal condition might be violated. \citet{2013Valori} have shown that, in addition to $E_\mathrm{p}$ and $E_\mathrm{f}$, the magnetic energy are contributed by three more terms in non-solenoidal field, which come from the non-solenoidal potential field, non-solenoidal current carrying field, and the mixture of the two fields. In such cases, the free magnetic energy derived from Equation~(\ref{eqn:Ef}) contains large uncertainties. In some cases, the calculated free magnetic energy is even negative. Such results are caused by the deviations from the solenoidal condition of the magnetic field. A good magnetic field model should have as small magnetic divergence as possible as demonstrated in \citet{2013Valori}.

The magnetic energy change in a volume $V$ is derived by the dot product between the magnetic field and the magnetic induction equation \citep{2006Schuck}:
\begin{equation}
\dfrac{\partial}{\partial t} \dfrac{B^2}{2\mu_0} =
\dfrac{1}{\mu_0} \nabla \cdot [(\mathbf{v} \times \mathbf{B})
\times \mathbf{B}] - \dfrac{1}{\mu_0} \nabla \cdot \left(
\dfrac{\mathbf{J} \times \mathbf{B}}{\sigma} \right) + (\mathbf{v}
\times \mathbf{B}) \cdot \mathbf{J} - \dfrac{J^2}{\sigma } .
\label{eqn:dBdt1}
\end{equation}
Integrating Equation~(\ref{eqn:dBdt1}) in the volume $V$, we derive the time evolution of the magnetic energy $E$:
\begin{equation}
\dfrac{\mathrm{d} E}{\mathrm{d} t} = \dfrac{1}{\mu_0}
\displaystyle\int\limits_{S} \left[ \mathbf{B} \times (\mathbf{v}
\times \mathbf{B}) + \dfrac{\mathbf{J} \times \mathbf{B}}{\sigma}
\right] \cdot \hat{\mathbf{n}} \; \mathrm{d} S -
\displaystyle\int\limits_{V} \left[ \mathbf{v} \cdot (\mathbf{J}
\times \mathbf{B}) + \dfrac{J^2}{\sigma} \right] \; \mathrm{d} V .
\label{eqn:dBdt2}
\end{equation}
The normal vector $\hat{\mathbf{n}}$ directs to the inner side of the volume. The energy change is caused by both a surface integration of energy injection flux and a volume integration of energy release density. The energy injection flux includes both the ideal process of the Poynting flux ($\mathbf{B} \times (\mathbf{v} \times \mathbf{B})/\mu_0$) and resistive process of the slippage of magnetic field lines in the plasma on the surface ($\mathbf{J} \times \mathbf{B}/(\sigma \mu_0$)). Similarly, the energy release density includes both the ideal process of the work done by the magnetic field on the plasma ($\mathbf{v} \cdot (\mathbf{J} \times \mathbf{B})$) and the resistive process of the Joule dissipation ($J^2/\sigma$). If the magnetic field is force-free ($\mathbf{J} \times
\mathbf{B} = 0$) and the plasma is ideal ($\sigma$ approaches infinity), the energy change can be computed by the Poynting flux through the photospheric surface $S_\mathrm{p}$:
\begin{equation}
\dfrac{\mathrm{d} E}{\mathrm{d} t} = \dfrac{1}{\mu_0}
\displaystyle\int\limits_{S_\mathrm{p}} \mathbf{B} \times
(\mathbf{v} \times \mathbf{B}) \cdot \hat{\mathbf{n}} \;
\mathrm{d} S . \label{eqn:dBdt3}
\end{equation}

\section{Magnetic Topology Analysis} \label{sec:topology}
The terminology ``topology'' is defined as the unchangeable geometrical properties that are preserved under smooth deformations. The topology of a magnetic field could be described by the magnetic field line linkages, i.e., the field line mapping. The singular topology skeletons \citep{1997Priest}, such as magnetic null points, spines, fans, and bald patches, are places where magnetic field line linkages are discontinuous. There are also places where the linkages are continuous but change drastically, which are named as quasi-separatrix layers \citep[QSLs;][]{1995Priest,1996Demoulin,2002Titov}. Magnetic topology is intimately related to magnetic reconnection, because the magnetic topology skeletons (null points, spines, fans, bald patches, and QSLs) divide topologically distinct magnetic domains, and electric current sheets are prone to be built on the interface between the domains by magnetic shear \citep{2006Aulanier}. Meanwhile, the conductivity in the solar atmosphere is high and the upper chromosphere and lower corona are even dominated by magnetic forces. Magnetic field lines are then frozen into plasma and they govern single particle motion, thermal conductivity, and Alfv\'en wave propagation \citep{2005Longcope}. Therefore, magnetic topology is responsible for the morphology of various manifestations of solar activities, such as location of hard X-ray sources, filament barb chirality, and flare ribbon shape and motion.

\subsection{Null Point and Spine-Fan Structures} \label{sec:null}
Magnetic null point is a place where $\mathbf{B}=0$. To study the structure of a magnetic field supposed to contain null points, it is necessary to locate their positions. In some special cases, it might be possible to solve the equation $\mathbf{B}=0$ analytically. For general cases such as a magnetic field model based on observations, searching for null points must be done by a numerical method. The magnetic field is discrete in numerical models, so it is usually assumed to be trilinear between neighboring grids to fill the 3D space. Pascal D\'emoulin developed an FORTRAN code\footnote{\url{http://www.lesia.obspm.fr/fromage/}} to search for null points using a modified Powell hybrid method. A Newton--Raphson method has also been employed to search for null points \citep{2007Haynes,2011Titov,2012Sun}. An alternative method to locate the positions of null points is based on the Poincar\'e index \citep{1992Greene,2005Zhao,2008Zhao}. \citet{2007Haynes} provided a comparison of the Newton--Raphson method and the Poincar\'e index method.

For the first order the magnetic structure at the vicinity of a magnetic null point can be described by the linear term of its Taylor expansion:
\begin{equation}
\mathbf{B} = \mathbf{M} \cdot \mathbf{r}, \label{eqn:null}
\end{equation}
where $\mathbf{M}$ is the Jacobian matrix with $M_{ij} = \partial B_i / \partial x_j$ $(i,j=1,2,3)$, and $\mathbf{r}$ is the position vector of the place of interest related to the null point. The matrix $\mathbf{M}$ can be diagonalized by solving the eigenfunction equation. There are three eigenvalues $\lambda_1, \lambda_2$, and $\lambda_3$ corresponding to three eigenvectors. Due to the solenoidal condition of the magnetic field, one can prove that $\lambda_1 + \lambda_2 + \lambda_3 = 0$. So there can be one negative eigenvalue (say, $\lambda_3$) and two positive ones ($\lambda_1$ and $\lambda_2$) or one positive eigenvalue ($\lambda_3$) and two negative ones ($\lambda_1$ and $\lambda_2$). The former represents a positive null point and the latter a negative null point. The single eigenvector associated with $\lambda_3$ determines the local direction of a spine, and the other two eigenvectors associated with $\lambda_1$ and $\lambda_2$ determine the local surface of a fan.

The three eigenvectors are not necessarily to be orthogonal, while they are always linearly independent. If there are electric currents perpendicular to the spine, the fan would be inclined to the spine \citep{1996Parnell}. In addition, \citet{1996Parnell} provided a thorough analysis of the linear structure of a magnetic null point. If the magnetic field is potential without any electric currents, matrix $\mathbf{M}$ is symmetric with three real eigenvalues. The fan is perpendicular to the spine. If there are electric currents in the vicinity of the null, there are four different types of spine-fan structures depending on the electric current parallel ($j_\parallel$) and perpendicular ($j_\perp$) to the spine. First, when $|j_\parallel|$ is less than or equal to a threshold of the electric current, $j_\mathrm{thresh}$, and $j_\perp=0$, there are three distinct or two equal real eigenvalues and the fan is perpendicular to the spine. Second, when $|j_\parallel| \le j_\mathrm{thresh}$ and $|j_\perp|>0$, the fan is inclined to the spine. Third, when $|j_\parallel| > j_\mathrm{thresh}$ and $j_\perp=0$, there are one real and two conjugate complex eigenvalues, and the fan is perpendicular to the spine. Finally, when $|j_\parallel| > j_\mathrm{thresh}$ and $|j_\perp|>0$, the fan is inclined to the spine.

If there are two null points interconnected with each other, the two fans intersect at a separator, which is a line connecting the two null points. A separator in 3D space is analogous to an X-point in 2D magnetic field. The spine of one null point could be the boundary line of a fan of another null point. The end of a spine is a magnetic source or at the infinite distance \citep{1997Priest}.

\subsection{Bald Patches and Magnetic Dips} \label{sec:bald}

Bald patches are locations on the photosphere and a polarity inversion line, where the magnetic field line is tangent to them and shaped concave up \citep{1993Titov,1996Bungey}. Seen from above, the local magnetic field transits from negative polarity to positive one, which forms an inverse polarity configuration. The magnetic field $\mathbf{B}$ at the bald patches satisfies
\begin{equation}
(\mathbf{B} \cdot \nabla) B_z > 0, \label{eqn:bald}
\end{equation}
where $B_z=0$. Similar to bald patches, magnetic dips also satisfy Equation~(\ref{eqn:bald}). The difference is that the latter are distributed in the whole space but not restricted to the photosphere.

\citet{1993Titov} showed that the magnetic field strength above bald patches and magnetic dips increases with height for force-free magnetic field. In general, the Lorentz force can be expanded as
\begin{equation}
\mathbf{J} \times \mathbf{B} = \frac{1}{\mu_0}(\mathbf{B} \cdot \nabla)\mathbf{B} - \nabla \left(\frac{B^2}{2\mu_0}\right). \label{eqn:lorentz}
\end{equation}
For a force-free field where $\mathbf{J} \times \mathbf{B} = 0$, the magnetic tension and the gradient of magnetic pressure balance each other as
\begin{equation}
\frac{B^2}{R} - \frac{\partial(B^2/2)}{\partial n} = 0, \label{eqn:tension}
\end{equation}
where $\mathbf{B}=B\mathbf{s}$ and $\mathbf{s}$ is a unit vector along $\mathbf{B}$. The normal unit vector $\mathbf{n}$ is defined as $\mathbf{n}/R = d\mathbf{s}/ds$, where $R$ is the radius of curvature. Therefore, for a magnetic field line that is concave up, one gets
\begin{equation}
\frac{1}{B} \frac{\partial B}{\partial z} = \frac{\mathbf{n} \cdot \mathbf{e}_z}{R} > 0, \label{eqn:bald2}
\end{equation}
which implies that the magnetic strength increases with height.

\subsection{Quasi-Separatrix Layers} \label{sec:qsl}
QSLs are 3D thin volumes across which the gradient of the magnetic field linkage is large. They divide a magnetic field into different magnetic domains, between which the field line linkage changes drastically. QSLs are a generalization of true separatrix, where the magnetic field linkage is discontinuous. QSLs have a finite thickness, unlike a separatrix that is infinitely thin; or, to say, a QSL is a 3D thin volume, while a separatrix is a 2D surface.

\citet{1996Demoulin} proposed a natural way to compute the locations of QSLs. In a magnetic field, one could integrate a magnetic field line from position $(x, y, z)$ to both directions with a distance $s$ on each side. The two end points $(x', y', z')$ and $(x'', y'', z'')$ define a vector $(X_1, X_2, X_3) = (x''-x', y''-y', z''-z')$. If the magnetic field mapping changes fast, a small change in the position $(x, y, z)$ would cause a large change in the vector $(X_1, X_2, X_3)$. A natural quantification of this change is the summation of the square of each element in the Jacobian matrix for this vector, namely, the norm
\begin{equation}
N(x,y,z,s) = \sqrt{\displaystyle\sum\limits_{i=1}^3 \left[
\left(\frac{\partial X_i}{\partial x} \right)^2 +
\left(\frac{\partial X_i}{\partial y} \right)^2 +
\left(\frac{\partial X_i}{\partial z} \right)^2 \right]} .
\label{eqn:norm1}
\end{equation}
The parameter $s$ is free and could be defined by a geometrical boundary or a wave propagation distance. The QSLs are places with $N \gg 1$. Equation~(\ref{eqn:norm1}) can be simplified if $z'=z''=0$ and the footpoints $(x', y', z')$ and $(x'', y'', z'')$ are assumed to be tied on the photosphere:
\begin{equation}
N_\pm \equiv N(x_\pm,y_\pm) = \sqrt{ \left(\frac{\partial
X_\mp}{\partial x_\pm} \right)^2 + \left(\frac{\partial
X_\mp}{\partial y_\pm} \right)^2 + \left(\frac{\partial
Y_\mp}{\partial x_\pm} \right)^2 + \left(\frac{\partial
Y_\mp}{\partial y_\pm} \right)^2 } , \label{eqn:norm2}
\end{equation}
where $(X_\mp, Y_\mp) = (x_\mp-x_\pm, y_\mp-y_\pm)$ \citep{1995Priest}.

\citet{2002Titov} found that $N_+$ and $N_-$ usually do not equal each other even along the same magnetic field line. They proposed a new parameter, which is uniform along a magnetic field line, the squashing degree $Q$, to measure the mapping of magnetic field lines
\begin{equation}
Q = \frac{N_+^2}{|\Delta_+|} = \frac{N_-^2}{|\Delta_-|}, \label{eqn:q1}
\end{equation}
where $\Delta_+$ and $\Delta_-$ are the determinants of the following two Jacobian matrices $\mathcal{D}_+$ and $\mathcal{D}_-$:
\begin{equation}
\mathcal{D}_+ = \left(
\renewcommand{\arraystretch}{2.0}
\begin{array}{l l}
\dfrac{\partial X_-}{\partial x_+} & \dfrac{\partial X_-}{\partial
y_+} \\
\dfrac{\partial Y_-}{\partial x_+} & \dfrac{\partial Y_-}{\partial
y_+}
\end{array} \right) ,
\end{equation}
and
\begin{equation}
\mathcal{D}_- = \left(
\renewcommand{\arraystretch}{2.0}
\begin{array}{l l}
\dfrac{\partial X_+}{\partial x_-} & \dfrac{\partial X_+}{\partial
y_-} \\
\dfrac{\partial Y_+}{\partial x_-} & \dfrac{\partial Y_+}{\partial
y_-}
\end{array} \right) .
\end{equation}
Denoting the normal magnetic field at the positive and negative polarities by $B_{n+}$ and $B_{n-}$, respectively, \citet{2002Titov} showed that
\begin{equation}
Q = \frac{N_+^2}{|B_{n+}/B_{n-}|} = \frac{N_-^2}{|B_{n-}/B_{n+}|}. \label{eqn:q2}
\end{equation}
The QSLs are places where $Q \gg 2$. Two QSLs may intersect at a place that is defined as a hyperbolic flux tube, where the squashing degree $Q$ is very large and magnetic reconnection is prone to occur.

\citet{2012Pariat} proposed a numerical scheme to compute the squashing degree $Q$ in a 3D volume. Following \citet{2012Pariat}, this method has been implemented by some other authors and applied to 3D magnetic topologies of active regions or eruptions \citep{2012Savcheva2,2012Savcheva1,2013Guo,2014Zhao,2016Zhao,2015Yang,2016Yang,2016Liu,2016Tassev}.

\subsection{Applications to Observations} \label{sec:application}
Magnetic topology analysis has been applied to interpret various solar activities and phenomena, such as solar flares, coronal jets, filament structures, and magnetic flux rope eruption. First, magnetic null points are closely related with the morphologies of solar flare ribbons. Flare ribbons are found to stop at the border of the closest large-scale QSL. Magnetic null points are also involved in magnetic structures responsible for flares with an EUV late phase. Secondly, coronal jets could occur in a magnetic null configuration or QSLs associated with bald patches. Thirdly, solar filaments are supported by magnetic dips. The chirality of filament barbs (right bearing or left bearing) can be determined by the magnetic helicity (positive or negative) and magnetic configuration (magnetic flux rope or magnetic arcade). Finally, magnetic flux rope, magnetic null points and QSLs may interact each other and lead to solar eruptions and magnetic reconnection.

\begin{figure}
\begin{center}
\includegraphics[width=1.0\textwidth]{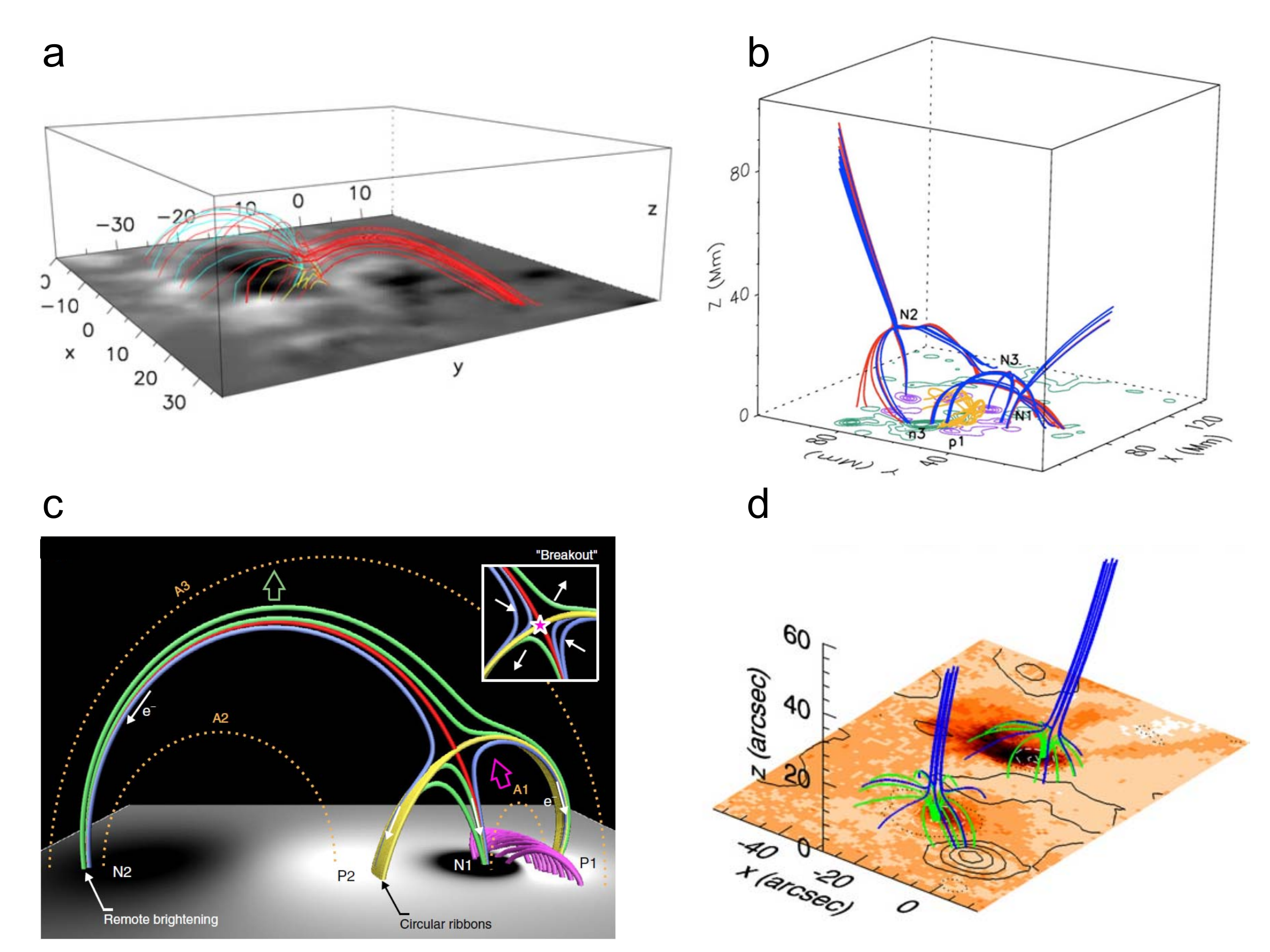}
\caption{Magnetic null points in different magnetic field models. \textbf{a.} Red, cyan, and yellow lines represent the magnetic field lines in the vicinity of a null point in a potential field model adapted from \citet{2009Masson} and reproduced by permission of the AAS. The grey-scale image shows the vertical magnetic field. \textbf{b.} Blue and red lines represent the magnetic field lines in the vicinities of three null points in a nonlinear force-free field. Yellow lines denote some sheared and twisted magnetic field lines. Contours show the vertical magnetic field. The figure is adapted from \citet{2014Mandrini} and reproduced by permission of Springer Science+Business Media Dordrecht. \textbf{c.} A toy model showing both the magnetic field lines close to the null point and sheared field lines under it. The figure is adapted from \citet{2013Sun} and reproduced by permission of the AAS. \textbf{d.} Blue and green lines denote the field lines close to two null points. Contours show the vertical magnetic field and the color image is a soft X-ray image observed by the X-ray Telescope aborad \textit{Hinode}. The figure is from \citet{2012Zhang} and reproduced by permission of the AAS.
} \label{fig:null}
\end{center}
\end{figure}

Circular ribbon flares are found to be associated with magnetic null points and spine-fan structures. A typical circular ribbon flare has three ribbons, a circular, an inner, and a remote one, which correspond to the traces of the fan separatrix, the inner spine, and the outer spine on the bottom, respectively. \citet{2009Masson} analyzed the magnetic topology (Figure~\ref{fig:null}a) and magnetic reconnection in such a circular ribbon flare. \citet{2012Reid} further studied the same flare and found some compact X-ray sources on the circular and inner ribbons and an extended source on the spine. They proposed that a hyperbolic flux tube embedded in the fan structure could explain the presence of co-spatial X-rays with the strongest UV emission. Note that there are other possibilities to explain such phenomena. For example, \citet{2015Yang} found a magnetic flux rope underlying a spine-fan structure. Compact X-ray sources could be explained by magnetic reconnection in and around the flux rope itself. Similar structures with a magnetic flux rope lying under three null points (Figure~\ref{fig:null}b) have also been found in \citet{2014Mandrini}. A special point is that the three null points in \citet{2014Mandrini} are highly asymmetric, where the fan separatrix is broken into two sections. In this event, the three flare ribbons correspond to traces of the two sections of the fan and the inner spine, while no flare ribbon is found at the footpoint of the remote spine.

Although magnetic reconnection is prone to occur in QSLs, when we use a force-free field to model the magnetic structure, we have to be cautious about the explanation. Flare ribbons are associated with the hyperbolic flux tubes beneath an erupting magnetic flux rope. However, the dynamics of flare ribbons cannot be fully modeled by a force-free field model. A well observed feature for typical two-ribbon flares is that the two ribbons often separate from each other, while QSLs calculated from a force-free field model has rarely significant change before and after a flare, especially for the traces on the bottom boundary. This is because flare ribbons are associated with QSLs developed in a dynamical process, which cannot be derived from force-free field models. Moreover, QSLs derived in a model are only possible places for magnetic reconnection. There exists QSLs that do not participate in magnetic reconnection. It has been found that flare ribbons tend to stop at the border of the closest large-scale QSLs, when the size of the QSLs is comparable to that of the active region hosting the eruption and the QSLs do not stride over the erupting magnetic flux rope \citep{2012Chen,2012Guo2}. On the other hand, without using the QSL method, \citet{2016Jiang2} directly reproduced the location of the flare ribbons of a major confined flare in active region 12192 by tracing the footpoints of the field lines from the reconnection current sheet in their data-driven MHD simulation.

The EUV late phase was discovered by the EUV Variability Experiment \citep[EVE;][]{2011Woods,2012Woods} on board \textit{SDO}. The key feature is that there appears a second peak in the relatively warm emission (e.g., Fe \uppercase\expandafter{\romannumeral15} 28.4~nm and Fe \uppercase\expandafter{\romannumeral16} 33.5~nm) after the main GOES soft X-ray peak. There should exist a second set of higher and longer coronal loops to produce the EUV late phase. Some typical flares with an EUV late phase have been analyzed in \citet{2012Hock}, \citet{2013Liu}, \citet{2013Dai}, \citet{2013Sun}, and \citet{2014Li}. Regarding the magnetic topology related to flares with an EUV late phase, \citet{2013Sun} found that a hot spine is a viable magnetic structure (Figure~\ref{fig:null}c). \citet{2014Li} found more cases where the spine is the key magnetic structure for producing the EUV late phase. However, in a specific case of the X2.1 flare on 2011 September, both a spine and some large-scale magnetic loops are found \citep{2014Li}. \citet{2013Dai} showed that the late phase emission is produced by the large-scale magnetic loops but not the spine.

\begin{figure}
\begin{center}
\includegraphics[width=1.0\textwidth]{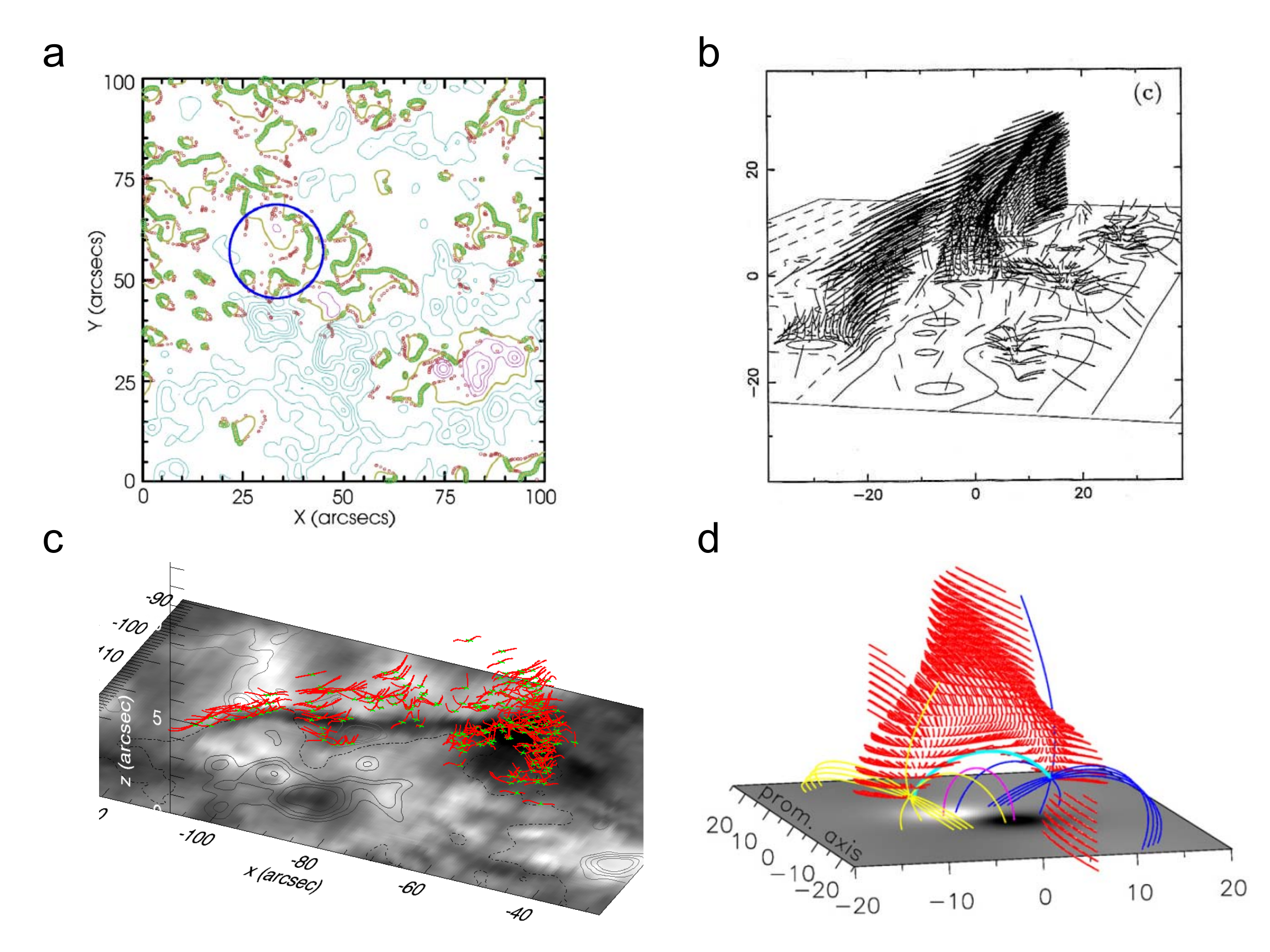}
\caption{Bald patches and magnetic dips in different magnetic field models. \textbf{a.} Green lines mark the positions of bald patches derived from a nonlinear force-free field model. The figure is adapted from \citet{2013Schmieder} and reproduced with permission from Astronomy \& Astrophysics, \copyright~ESO. \textbf{b.} Dark line sections denote magnetic field line sections with magnetic dips that are computed in a linear force-free field model. The figure is adapted from \citet{1998Aulanier2} and reproduced with permission from Astronomy \& Astrophysics, \copyright~ESO. \textbf{c.} Green crosses denote the positions of magnetic dips computed in a nonlinear force-free field model. Red line sections represent the magnetic field line sections passing the magnetic dips. The figure is adapted from \citet{2010Guo2} and reproduced by permission of the AAS. \textbf{d.} Red line sections denote magnetic field line sections with magnetic dips in a linear force-free field model. Other lines are magnetic field lines associated with two null points and an underlying magnetic arcade. The figure is adapted from \citet{2012Dudik} and reproduced by permission of the AAS.
} \label{fig:dip}
\end{center}
\end{figure}

There are two categories of 2D models for coronal jets, namely, the emerging flux model \citep{1977Heyvaerts,1992Shibata265} and the converging flux model \citep{1994Priest}. In 3D, both the magnetic null point model \citep{2009Pariat,2010Pariat,2012Zhang} and bald patch model \citep{2013Guo2,2013Schmieder} have been proposed. Many coronal jets have a periodicity of about tens of minutes to several hours. \citet{2010Pariat} proposed that photospheric twisting motion drives the null point and spine-fan structure to reconnect periodically. \citet{2012Zhang} proposed that the modulation of trapped slow-mode waves along the spine field lines (Figure~\ref{fig:null}d) might cause such a periodicity. \citet{2013Guo2} found that magnetic reconnection in QSLs associated with bald patches could cause the eruption of coronal jets. \citet{2013Schmieder} discovered that coronal jets might appear in a bundle of twisted field lines, among which at least one of the footpoints is not connected to the bottom. Similar to \citet{2013Guo2}, magnetic reconnection also occurs in the QSLs associated with bald patches (Figure~\ref{fig:dip}a). Coronal jets could gain twists from the field lines during the expulsion of the jet plasma.

\citet{1957Kippenhahn} and \citet{1974Kuperus} proposed two magnetic field configurations, one with a normal polarity (where the horizontal magnetic field has a component pointing from the positive to negative polarity) and the other with an inverse polarity (opposite to the normal one), respectively, to explain the magnetic structure for filaments. Both magnetic configurations contain magnetic dips. \citet{1998Aulanier1} proposed a series of linear force-free models to explain the presence of filament barbs, where magnetic dips are assumed to support filament material. \citet{1998Aulanier2} found that the shape of a filament is determined by the distribution of magnetic dips (Figure~\ref{fig:dip}b) in a linear magnetic field. \citet{2010Guo2} also found that magnetic dips resemble well the shape of a filament observed in H$\alpha$ using a nonlinear force-free field model (Figure~\ref{fig:dip}c). Additionally, both a magnetic flux rope and sheared magnetic arcades are found in the same filament. The chirality of filament barbs are determined by the magnetic helicity and magnetic configuration simultaneously. In a magnetic field with negative magnetic helicity, a magnetic flux rope (with inverse polarity) induces right bearing filament barbs, and sheared arcades (with normal polarity) induce left bearing filament barbs. Similarly, in a magnetic field with positive magnetic helicity, a magnetic flux rope induces left bearing filament barbs, and sheared arcades induce right bearing ones. This result has also been discussed in \citet{2014Chen}. Figure~\ref{fig:dip}d shows a linear force-free field model for explaining polar crown prominences with bubbles and plumes \citep{2012Dudik}, which are interpreted as a separator magnetic reconnection rather than the Rayleigh--Taylor instability \citep{2011Hillier,2012Hillier1,2012Hillier2}.

\begin{figure}
\begin{center}
\includegraphics[width=1.0\textwidth]{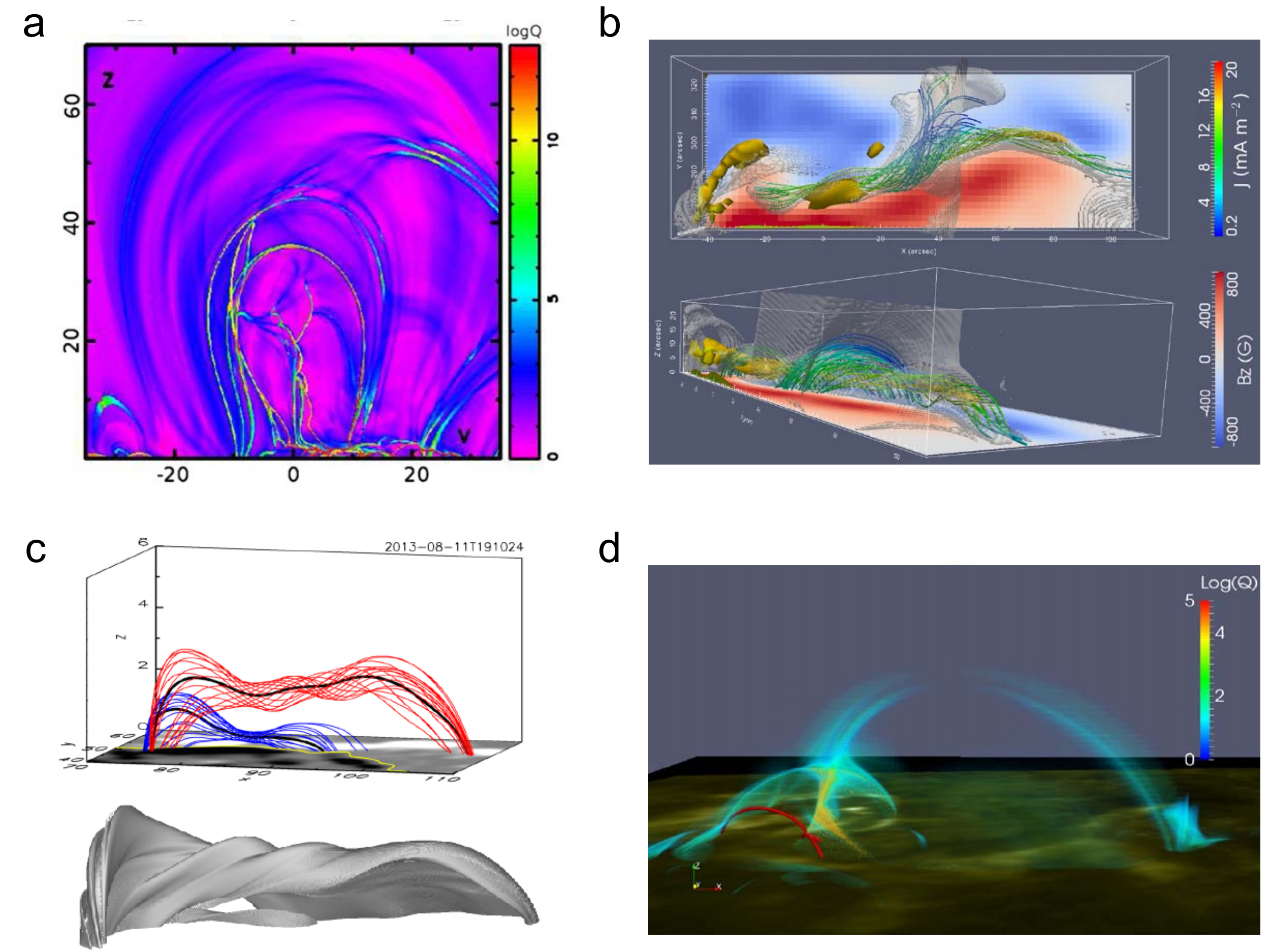}
\caption{QSLs in different magnetic field models. \textbf{a.} Distribution of the logarithm of the squashing degree $Q$ for a nonlinear force-free field model from \citet{2012Savcheva2} and reproduced by permission of the AAS. \textbf{b.} White semitransparent surfaces represent the isosurface of the squashing degree at $Q=10^4$. Solid lines are selected magnetic field lines of a nonlinear force-free field model. Yellow surfaces are isosurface of the electric current density. The image on the bottom shows vertical magnetic field. The figure is adapted from \citet{2013Guo} and reproduced by permission of the AAS. \textbf{c.} Red and blue lines show the twisted magnetic field lines of a nonlinear force-free field model. Black lines denote the axes of the magnetic flux rope. The lower image shows the isosurface of the twist number that equals $-1$. The isosurface of the twist number has a similar distribution as that of the squashing degree. The figure is from \citet{2016Liu} and reproduced by permission of the AAS. \textbf{d.} The semitransparent surfaces represent the 3D distribution of the logarithm of the squashing degree $Q$ computed from a potential field model. The red lines delineate some field lines from a magnetic null point. The image on the bottom is a \textit{SDO}/AIA 1600~\AA \ image. The figure is from \citet{2015Yang} and reproduced by permission of the AAS.
} \label{fig:qsl}
\end{center}
\end{figure}

\citet{2012Savcheva1} studied the QSLs of nonlinear force-free field models for a long-lasting coronal sigmoid, which is a signature of a magnetic flux rope. The sigmoid keeps stable for several days although there are bald patches in the constructed magnetic field. It erupts as long as the magnetic field exhibits hyperbolic flux tubes. \citet{2012Savcheva2} further compared the QSLs computed from both the nonlinear force-free field model (Figure~\ref{fig:qsl}a) and the MHD simulation, and suggested that magnetic reconnection at the hyperbolic flux tube under the flux rope and torus instability jointly cause the observed CME. \citet{2013Guo} studied the QSLs, twist accumulation, and magnetic helicity injection of a magnetic flux rope before a major flare and CME. It is found that the flux rope is surrounded by QSLs (Figure~\ref{fig:qsl}b). The twist and magnetic helicity are injected into the flux rope by continuous magnetic reconnection in the QSLs. The magnetic topology structure, evolution, and stability of a double-decker flux rope (Figure~\ref{fig:qsl}c) have also been studied by \citet{2016Liu}. \citet{2015Yang} presented a 3D QSL with both a spine-fan separatrix and a large-scale quadrupolar structure (Figure~\ref{fig:qsl}d). The 3D QSL resembles very well the EUV emission in 94~\AA \ observed by \textit{SDO}/AIA. A magnetic flux rope is found below the fan. The interplay between MHD instability and magnetic reconnection in the QSL structures could explain the eruption process revealed in multi-wavelength observations. More applications of QSLs to the interpretation of observations can be found in \citet{2014Zhao} and \citet{2016Janvier}.

\section{Magnetic Helicity Computation} \label{sec:helicity}
Magnetic helicity, the volume integration of the product of the vector potential and the magnetic field, is not only conserved in ideal MHD process but also approximately conserved in resistive process with magnetic reconnection \citep{1974Taylor,1984Berger2}. The conservation of magnetic helicity constrains the final state in the relaxation process of plasma confinement \citep{1958Woltjer,1974Taylor}. Magnetic helicity is preferentially negative in the northern solar hemisphere and positive in the southern one \citep{1994Rust,1994Pevtsov,1995Pevtsov,2003Pevtsov,1997Zirker,1998Zhang,1999Zhang,1998Bao,2017Ouyang} and it seems unchanged with solar cycle, although this point is under debate \citep{2002Hagino,2004Hagino,2010Zhang}. Magnetic helicity also plays a critical role in the eruption mechanisms of CMEs \citep{2003Zhang,2006Zhang}, the formation of filament channels \citep{2013Antiochos,2015Knizhnik,2017Knizhnik}, and the physical process in magnetic field dynamos \citep{2005Brandenburg,2006ZhangHQ}.

Magnetic helicity measures the topological complexity of a bundle of magnetic field lines. In a volume $V$, magnetic helicity $H_m$ is expressed as
\begin{equation}
H_m = \displaystyle\int\limits_{V} \mathbf{A} \cdot \mathbf{B} \;
\mathrm{d} V , \label{eqn:helic}
\end{equation}
where $\mathbf{A}$ denotes the vector potential and $\mathbf{B} = \nabla \times \mathbf{A}$. Note that Equation~(\ref{eqn:helic}) is gauge invariant only when the magnetic field is closed in a volume with no normal component on the boundary. In realistic cases, such as a magnetic field rooted on the photosphere, the field is open with fluxes passing through the boundaries. \citet{1984Berger} and \citet{1985Finn} proposed a relative magnetic helicity, $H$, which is gauge invariant for both closed and open configurations:
\begin{equation}
H = \displaystyle\int\limits_{V} (\mathbf{A} +
\mathbf{A}_\mathrm{p}) \cdot (\mathbf{B} - \mathbf{B}_\mathrm{p})
\; \mathrm{d} V , \label{eqn:helic2}
\end{equation}
where $\mathbf{B}_\mathrm{p}$ is the reference field with $\nabla \times \mathbf{A}_\mathrm{p} = \mathbf{B}_\mathrm{p}$. The reference field is usually selected as the potential field and it has the same normal component as the magnetic field $\mathbf{B}$ on the boundary $\mathbf{S}$ so that
\begin{equation}
(\nabla \times \mathbf{A}_\mathrm{p}) \cdot \hat{\mathbf{n}} = \mathbf{B} \cdot \hat{\mathbf{n}}, \label{eqn:bound1}
\end{equation}
and
\begin{equation}
(\nabla \times \mathbf{A}) \cdot \hat{\mathbf{n}} = \mathbf{B} \cdot \hat{\mathbf{n}}. \label{eqn:bound2}
\end{equation}
The unit vector $\hat{\mathbf{n}}$ is normal to $\mathbf{S}$ and directs to the inner side of the volume.

Although the relative helicity Equation~(\ref{eqn:helic2}) is widely used in computations of magnetic helicity in open configurations, there are other expressions and interpretations of the magnetic helicity. \citet{2011Low} pointed out that the relative magnetic helicities might be not conserved due to the evolution of the reference field itself. \citet{2006Low} and \citet{2011Low} proposed to compute the ``Lagrangian helicity'' and ``absolute helicity'' using a two flux representation or the representation of \citet{1957Chandrasekhar}. \citet{2014Prior} argued that although the relative helicity Equation~(\ref{eqn:helic2}) is gauge invariant, it is dependent on the choice of the reference field. \citet{2014Prior} proposed to use the classical magnetic helicity Equation~(\ref{eqn:helic}) but fix the gauge as the winding gauge (Equation (18) in \citealt{2014Prior}). Other gauges can be found in \citet{1984Jensen} and \citet{2006Hornig}. The value and interpretation of the magnetic helicity are dependent on the choice of the expressions (classical, relative, Lagrangian, or absolute magnetic helicity) and gauges (Coulomb gauge, DeVore gauge, winding gauge, and so on). It still needs further studies to clarify these issues.

In the following, we introduce three practical methods to compute the magnetic helicity, which include the finite volume method \citep{2011Rudenko,2011Thalmann,2012Valori,2013Yang,2014Rudenko,2014Moraitis}, the discrete flux tube method \citep{2010Guo,2013Guo,2012Georgoulis}, and the helicity flux integration method \citep{2001Chae,2005Pariat,2012Liu}. A series of papers have been devoted to benchmark, compare, and apply these methods \citep{2016Valori,2017Pariat,2017Guo}. Besides the above methods, \citet{2015Russell} proposed a field-line helicity method, where the helicity density is assigned to each individual field line. The helicity density in each field line is the integration of the vector potential along it. \citet{2008Longcope} proposed two generalizations of the relative magnetic helicity, one is the unconfined self-helicity and the other is the additive self-helicity, using different reference fields and considering sub-volumes in the corona. Further details can be found in the respective references.

\subsection{Finite Volume Method} \label{sec:fv}
The finite volume method integrates Equation~(\ref{eqn:helic2}) numerically with different gauges and boundary conditions. It requires an input of the full 3D magnetic field information, which is provided by theoretical force-free field models, numerical nonlinear force-free field models, and MHD models as discussed in Section~\ref{sec:extrapolation}. The output is the relative magnetic helicity at a certain moment. \citet{2016Valori} provided a detailed description of six implementations of the finite volume method, and made a thorough benchmark and comparison of the six algorithms regarding their accuracy, mutual consistency, and sensitivity.

The six algorithms for the finite volume method are divided into two categories according to the employed gauges, one being the Coulomb gauge ($\nabla \cdot \mathbf{A} = 0$) and the other being the DeVore gauge ($A_z=0$, \citealt{2000DeVore}). In the Coulomb gauge, the key problem is to solve the Laplace problem for the vector potential $\mathbf{A}_\mathrm{p}$:
\begin{equation}
\nabla^2 \mathbf{A}_\mathrm{p} = 0, \label{eqn:laplace}
\end{equation}
and the Poisson problem for $\mathbf{A}$:
\begin{equation}
\nabla^2 \mathbf{A} = -\mathbf{J}, \label{eqn:poisson}
\end{equation}
with the boundary conditions of Equations~(\ref{eqn:bound1}) and (\ref{eqn:bound2}) and the gauges $\nabla \cdot \mathbf{A} = 0$ and $\nabla \cdot \mathbf{A}_\mathrm{p} = 0$. The finite volume method in the Coulomb gauge has been implemented by \citet{2011Rudenko}, \citet{2014Rudenko}, \citet{2011Thalmann}, and \citet{2013Yang}.

\citet{2012Valori} derived the formula for the vector potential $\mathbf{A}$ using the DeVore gauge in a volume bounded by $[x_1, x_2]$, $[y_1, y_2]$, and $[z_1, z_2]$:
\begin{equation}
\mathbf{A} = \mathbf{b} + \hat{\mathbf{z}} \times \displaystyle\int\limits_{z}^{z_2} \mathbf{B} \mathrm{d}z' , \label{eqn:devore}
\end{equation}
with
\begin{equation}
\renewcommand{\arraystretch}{1.5}
\begin{array}{l}
b_x = -\displaystyle\frac{1}{2} \displaystyle\int\limits_{y_1}^{y} B_z(x,y',z_2) \mathrm{d} y', \\
b_y = \displaystyle\frac{1}{2} \displaystyle\int\limits_{x_1}^{x} B_z(x',y,z_2) \mathrm{d} x', \\
b_z = 0.
\end{array} \label{eqn:gauge}
\end{equation}
The vector potential $\mathbf{A}_\mathrm{p}$ is derived similarly by integrating $\mathbf{B}_\mathrm{p}$, which is solved by a scalar Laplace equation
\begin{equation}
\nabla^2 \Psi = 0, \label{eqn:pot_laplace}
\end{equation}
where $\mathbf{B}_\mathrm{p}=\nabla\Psi$. $\mathbf{B}_\mathrm{p}$ has the same normal component on the boundary as $\mathbf{B}$. The finite volume method in the DeVore gauge has been further implemented by \citet{2014Moraitis} and S. Anfinogentov \citep{2016Valori} following \citet{2012Valori} with only minor differences.

\subsection{Discrete Flux Tube Method} \label{sec:dt}
In the Coulomb gauge, it is found that Equation~(\ref{eqn:helic}) has a similar form to the Gauss linking number \citep{1969Moffatt,1984Berger,2006Berger}:
\begin{equation}
\mathcal{L}_k = \frac{1}{4\pi} \oint\limits_{\mathbf{x}} \oint\limits_{\mathbf{y}} \hat{\mathbf{T}}_\mathbf{x}(s) \times \hat{\mathbf{T}}_\mathbf{y}(s') \cdot \frac{\mathbf{r}}{|\mathbf{r}|^3} \mathrm{d}s' \, \mathrm{d}s , \label{eqn:linking}
\end{equation}
where $\mathbf{x}(s)$ and $\mathbf{y}(s')$ are the positions of two curves parameterized by $s$ and $s'$, $\hat{\mathbf{T}}_\mathbf{x}(s)$ and $\hat{\mathbf{T}}_\mathbf{y}(s')$ are the unit tangent vectors to $\mathbf{x}(s)$ and $\mathbf{y}(s')$, and $\mathbf{r}=\mathbf{x}(s) - \mathbf{y}(s')$. Note that Equation~(\ref{eqn:linking}) is valid for closed curves, and $\mathcal{L}_k$ is an integer and invariant to continuous deformations without tearing or gluing. The Gauss linking number can be expressed as the sum of the writhe $\mathcal{W}_r$ and twist $\mathcal{T}_w$, namely,
\begin{equation}
\mathcal{L}_k = \mathcal{W}_r + \mathcal{T}_w, \label{eqn:calug}
\end{equation}
which is known as the C$\mathrm{\check{a}}$lug$\mathrm{\check{a}}$reanu theorem \citep[e.g.,][]{1978Fuller,1992Moffatt,2006Berger}. The writhe measures the non-planarity of an axis curve alone:
\begin{equation}
\mathcal{W}_r = \frac{1}{4\pi} \oint\limits_{\mathbf{x}} \oint\limits_{\mathbf{x}} \hat{\mathbf{T}}(s) \times \hat{\mathbf{T}}(s') \cdot \frac{\mathbf{r}}{|\mathbf{r}|^3} \mathrm{d}s' \, \mathrm{d}s , \label{eqn:writhe}
\end{equation}
with $\mathbf{r}=\mathbf{x}(s) - \mathbf{x}(s')$. The twist measures the rotation of a secondary curve about the axis:
\begin{equation}
\mathcal{T}_w = \frac{1}{2\pi} \oint\limits_{\mathbf{x}} \hat{\mathbf{T}}(s)\cdot \hat{\mathbf{V}}(s) \times \frac{\mathrm{d}\hat{\mathbf{V}}(s)}{\mathrm{d}s} \mathrm{d}s , \label{eqn:twist}
\end{equation}
where $\hat{\mathbf{V}}(s)$ is a unit vector normal to $\hat{\mathbf{T}}(s)$ and pointing from the axis to the secondary curve. \citet{1978Fuller} pointed out that each of the three quantities, $\mathcal{L}_k$, $\mathcal{W}_r$, and $\mathcal{T}_w$, has a unique property that is not possessed by the other two. Specifically, $\mathcal{L}_k$ is topologically invariant, $\mathcal{W}_r$ depends only on the axis, and $\mathcal{T}_w$ is additive and quantified by a local density.

With the above analysis, \citet{1984Berger} assigned the magnetic helicity with a geometrical (or, topological) meaning under the Coulomb gauge. If a magnetic field is divided into a finite number ($N$) of flux tubes, the magnetic helicity arises both from the internal structure, namely twist and writhe, of each magnetic flux tube and the linkage and knotting of different flux tubes. The former is known as the self helicity, and the latter mutual helicity. Quantitatively, \citet{1984Berger} found that for a closed magnetic configuration (also refer to \citealt{2006Demoulin}):
\begin{equation}
H_m \approx \sum_{i=1}^N \mathcal{L}_i \Phi_i^2 + \sum_{i=1}^N \sum_{j=1,j \ne i}^N \mathcal{L}_{i,j} \Phi_i \Phi_j \, , \label{eqn:helic_DT}
\end{equation}
where $\mathcal{L}_i$ includes the contributions from both the writhe and twist as $\mathcal{L}_i = \mathcal{W}_i + \mathcal{T}_i$, $\mathcal{L}_{i,j}$ is the linking number between flux tubes $i$ and $j$, and $\Phi_i$ and $\Phi_j$ denotes the magnetic fluxes of tubes $i$ and $j$, respectively. Note that the self helicity arises because the number $N$ of flux tubes is finite so that each tube contains a certain flux. If the number $N$ approaches infinity, the self helicity vanishes and only the mutual helicity exists \citep{1984Berger,2006Demoulin}.

The above analyses are valid for closed magnetic configurations, where no magnetic flux penetrates the boundary. For open configurations, Equations~(\ref{eqn:linking}) and (\ref{eqn:writhe}) are not applicable any more. However, Equation~(\ref{eqn:twist}) is still applicable with the integration only along an open curve. Furthermore, Equation~(\ref{eqn:helic_DT}) still holds true, but the interpretation has to be changed. In open magnetic configurations, the relative helicity is adopted and the linking number is redefined in \citet{2006Demoulin} and the writhe in \citet{2006Berger}. Especially, \citet{2006Demoulin} used the concept of helicity injection to define the linking number of open curves, and they proposed an internal angle method to quantify this number.

In practice, Equation~(\ref{eqn:helic_DT}) has been adopted for the computation of magnetic helicity with different implementations and assumptions. The twist number method \citep{2010Guo,2013Guo,2017Guo,2014Xia,2016Valori,2016Yang} only computes the magnetic helicity contributed by the twist. It is applicable for cases where only one major electric current channel exists and the writhe is negligible. The twist number method uses QSLs as introduced in Section~\ref{sec:qsl} to determine the boundary of a magnetic flux rope. QSLs have a property of being a magnetic flux surface. Therefore, it is meaningful to compute the magnetic helicity for only a magnetic flux rope. The connectivity-based method \citep{2007Georgoulis,2012Georgoulis,2012Tziotziou,2013Tziotziou,2014Tziotziou,2014Moraitis} computes the linking number using the internal angle method. This method can use either a magnetic connectivity provided by force-free, MHS, or MHD models, or a connectivity inferred by a minimal connection length.

\subsection{Helicity Flux Integration Method} \label{sec:fi}

Both the finite volume method and the discrete flux tube method require a 3D magnetic field for the computation of the magnetic helicity of a volume. However, the accumulated magnetic helicity in a volume can also be estimated from the integration of the helicity flux through a boundary surface, say, the bottom surface that usually has the largest helicity flux. \citet{1984Berger} derived the time variation of $H$ (also refer to \citealt{2015Pariat}):
\begin{equation}
\dfrac{\mathrm{d} H}{\mathrm{d} t} = -2
\displaystyle\int\limits_{S} (\mathbf{A}_\mathrm{p} \times
\mathbf{E}) \; \mathrm{d} \mathbf{S} - 2
\displaystyle\int\limits_{V} \mathbf{E} \cdot \mathbf{B} \;
\mathrm{d} V + 2 \displaystyle\int\limits_{S} \dfrac{\partial
\Psi}{\partial t} \mathbf{A}_\mathrm{p} \cdot \; \mathrm{d}
\mathbf{S} , \label{eqn:dHdt1}
\end{equation}
with $\mathbf{B}_\mathrm{p}=\nabla\Psi$. The boundary condition and gauge for the vector potential $\mathbf{A}_\mathrm{p}$ has a certain freedom. To simplify the expression, a particular boundary condition and gauge are selected, namely, the vector potential $\mathbf{A}_\mathrm{p}$ has no normal component on the boundary $\mathbf{S}$ and $\mathbf{A}_\mathrm{p}$ is solenoidal in the volume $V$:
\begin{equation}
\renewcommand{\arraystretch}{1.5}
\begin{array}{l}
\mathbf{A}_\mathrm{p} \cdot \hat{\mathbf{n}} = 0 , \\
\nabla \cdot \mathbf{A}_\mathrm{p} = 0.
\end{array} \label{eqn:gauge}
\end{equation}
Note that $\hat{\mathbf{n}}$ is defined as directing to the inner side of the volume. In some studies \citep[e.g.,][]{2014Pevtsov,2015Pariat}, it can also be defined as directing to the outer side of the volume, where the surface integration in Equation~(\ref{eqn:dHdt1}) should change its sign. Substituting Equation~(\ref{eqn:gauge}) into Equation~(\ref{eqn:dHdt1}) and using the Ohm's law
\begin{equation}
\mathbf{E} = -\mathbf{v} \times \mathbf{B}+\frac{\mathbf{J}}{\sigma}, \label{eqn:ohm}
\end{equation}
Equation~(\ref{eqn:dHdt1}) is reduced to
\begin{equation}
\dfrac{\mathrm{d}H}{\mathrm{d}t} = 2 \displaystyle\int\limits_{S}
[ (\mathbf{A}_\mathrm{p} \cdot \mathbf{B}) \mathbf{v} -
(\mathbf{A}_\mathrm{p} \cdot \mathbf{v}) \mathbf{B}] \cdot
\hat{\mathbf{n}} \; \mathrm{d} S - 2 \displaystyle\int\limits_{S}
\dfrac{1}{\sigma}(\mathbf{A}_\mathrm{p} \times \mathbf{J}) \cdot
\hat{\mathbf{n}} \; \mathrm{d} S - 2 \displaystyle\int\limits_{V}
\dfrac{1}{\sigma} \mathbf{J} \cdot \mathbf{B} \; \mathrm{d} V .
\label{eqn:dHdt2}
\end{equation}
Equation~(\ref{eqn:dHdt2}) shows that the helicity in a volume changes owing to helicity flux across the boundary and dissipation in the volume.

In the solar atmosphere, the conductivity can be regarded as infinitely large. So, the last two terms in Equation~(\ref{eqn:dHdt2}) are omitted. Then we have
\begin{equation}
\dfrac{\mathrm{d} H}{\mathrm{d} t} = -2
\displaystyle\int\limits_{S_\mathrm{p}} (\mathbf{A}_\mathrm{p}
\cdot \mathbf{u}) B_n \; \mathrm{d} S ,
\label{eqn:dHdt3}
\end{equation}
where only the flux across the photosphere $S_\mathrm{p}$ is considered and
\begin{equation}
\mathbf{u} = \mathbf{v}_t - \dfrac{v_n}{B_n} \mathbf{B}_t .
\label{eqn:ftvel2}
\end{equation}
\citet{2003Demoulin} argued that horizontal velocities derived from optical flow techniques as discussed in Section~\ref{sec:velocity} represent $\mathbf{u}$ in Equation~(\ref{eqn:ftvel2}) rather than $\mathbf{v}_t$. However, this viewpoint has been questioned by, for example \citet{2007Welsch} and \citet{2008Schuck}. The integrand in Equation~(\ref{eqn:dHdt3}) is defined as the helicity flux density,
\begin{equation}
G_A(\mathbf{x}) = -2 (\mathbf{A}_\mathrm{p} \cdot \mathbf{u}) B_n
. \label{eqn:ga}
\end{equation}
$G_A(\mathbf{x})$ has been adopted to compute the magnetic helicity flux distribution in active regions \citep{2001Chae,2002Kusano,2002Moon,2002Nindos}.

\citet{2005Pariat} proposed an alternative expression for the magnetic helicity flux
\begin{equation}
\dfrac{\mathrm{d} H}{\mathrm{d} t} = -\dfrac{1}{2\pi}
\displaystyle\int\limits_{S_\mathrm{p}}
\displaystyle\int\limits_{S'_\mathrm{p}} \dfrac{\mathrm{d}
\theta(\mathbf{r})}{\mathrm{d} t} B_n B'_n \; \mathrm{d} S' \;
\mathrm{d} S, \label{eqn:dHdt4}
\end{equation}
with
\begin{equation}
\dfrac{\mathrm{d} \theta(\mathbf{r})}{\mathrm{d} t} =
\dfrac{1}{r^2} \left(\mathbf{r} \times \dfrac{\mathrm{d}
\mathbf{r}}{\mathrm{d} t} \right)_n = \dfrac{1}{r^2} [\mathbf{r}
\times ( \mathbf{u} - \mathbf{u'})]_n ,
\label{eqn:dthdt}
\end{equation}
where $\mathbf{r} = \mathbf{x} - \mathbf{x'}$ is the position vector. Equation~(\ref{eqn:dHdt4}) shows that the magnetic helicity is injected by the rotation of each pair of elementary flux tubes weighted by their magnetic fluxes. Consequently, the helicity flux density $G_\theta(\mathbf{x})$ is defined as
\begin{equation}
G_\theta(\mathbf{x}) = -\dfrac{B_n}{2\pi}
\displaystyle\int\limits_{S'_\mathrm{p}} \dfrac{\mathrm{d}
\theta(\mathbf{r})}{\mathrm{d} t} B'_n \; \mathrm{d} S' . \label{eqn:gth}
\end{equation}
Neither $G_A(\mathbf{x})$ nor $G_\theta(\mathbf{x})$ itself is meaningful as a measurement of the helicity flux density. \citet{2005Pariat} proposed to use a proxy $G_\Phi(\mathbf{x})$, which meaningfully measures the connectivity-based helicity flux density per elementary magnetic flux tube. The method has been implemented by \citet{2014Dalmasse} and applied to observations by \citet{2013Dalmasse}.

\subsection{Applications to Models and Observations} \label{sec:app_hel}

\citet{2012Valori} implemented a finite volume method using the DeVore gauge and applied it to the force-free Titov-D\'emoulin model. They found that this method only needs a small volume to derive the full helicity content. \citet{2013Yang} applied another finite volume method using the Coulomb gauge to a data-driven MHD simulation, and found that the accumulated magnetic helicity in the volume coincides with the helicity flux injected through the boundaries. \citet{2015Pariat} also made a thorough test on the magnetic helicity conservation by computing the relative magnetic helicity in a finite volume and the integration of the helicity flux through the boundaries. It is found that the dissipation of the magnetic helicity is almost zero in a quasi-ideal MHD process, and the dissipation in a resistive process is very low ($< 2.2\%$). By contrast, the dissipation rate of magnetic energy is much higher, or tens of times the dissipation rate of magnetic helicity. \citet{2017Pariat2} proposed that the relative magnetic helicity can be used as a diagnostic tool for solar eruptivity. They found that the ratio of the magnetic helicity carried purely by the electric currents to the total relative helicity is very high for eruptive simulations. Thus, this parameter can be used to distinguish between eruptive and confined eruptions.

\citet{2013Guo} adopted the twist number method and helicity flux integration method to study the twist accumulation and magnetic helicity injection, respectively. It is found that only a small fraction of the injected helicity is transferred to the internal helicity of a magnetic flux rope. \citet{2016Valori} and \citet{2017Guo} found that the magnetic helicity contributed by the twist matches well the magnetic helicity contributed only by the electric current, which is gauge invariant representing part of the relative magnetic helicity. With the connectivity-based method, \citet{2012Tziotziou} calculated the free magnetic energy and relative magnetic helicity for 162 vector magnetic fields in 42 active regions, and found a clear linear correlation between the magnetic energy and helicity. Eruptive active regions possess both large free energy and relative helicity exceeding $4 \times 10^{31}$ erg and $2 \times 10^{42}$ Mx$^2$, respectively. \citet{2013Tziotziou} applied the connectivity-based method to a time series of 600 vector magnetic fields in active region 11158 and found that both the free magnetic energy and relative magnetic helicity are accumulated to sufficient amounts to power a series of solar eruptions.

\citet{2013Dalmasse} first applied a connectivity-based helicity flux density to vector magnetic fields observed by \textit{SDO}/HMI in active region 11158, and confirmed that the helicity flux density is mixed with both positive and negative signs. \citet{2014Liu} used the helicity flux integration method to study the helicity injection in emerging active regions. They found that about $61\%$ of the 28 emerging active regions follow the hemispheric rule, which states that the magnetic helicity is negative in the northern hemisphere and positive in the southern one. The helicity flux integration method has been widely used in various studies on the magnetic helicity injection through the boundaries \citep[e.g.,][]{2008Tian,2009Yamamoto,2009Yang,2010Chandra,2010Park,2011Zuccarello,2012Jing,2012Vemareddy,2014Romano,2016Lim}.

\section{Summary and Discussion} \label{sec:conclusion}
In this review, we introduce the recent progresses in observations, theories, and numerical methods related to magnetic field structures and dynamics in the solar atmosphere. The magnetic field on the photosphere is observed with the polarized light emergent from there, which is represented by the Stokes parameters. The magnetic field is derived by the inversion of the Stokes spectral profiles under certain assumptions for the physical environment of the photosphere. The transverse components of a vector magnetic field have an intrinsic 180$^\circ$ ambiguity, which has to be removed under additional physical assumptions. Correction of the projection effect is also necessary if the field of view is large or close to the solar limb. With a time series of vector magnetic fields, the velocity can be derived by the optical flow techniques.

There are different ways to obtain the 3D magnetic field and its evolution. We introduce some theoretical force-free field models, numerical nonlinear force-free field models, MHS models, and MHD models that are widely used. We also introduce some methods to preprocess the vector magnetic field observed on the photosphere as a suitable boundary condition for nonlinear force-free field extrapolation. Magnetic energy computation in a volume and from the boundaries is briefly discussed. To quantify the structure and stability of a magnetic field, the magnetic topology analysis and magnetic helicity computation are essential. The methods to pinpoint null points, bald patches, and QSLs are introduced. We also mention some applications of these concepts to the interpretation of observations. Some practical methods to compute the magnetic helicity are presented. They include the finite volume method, discrete flux tube method, helicity flux integration method, and other methods. Applications of magnetic helicity to interpreting solar eruptive activities are also mentioned.

It should be pointed out that most of the methods, including those described above and some others not mentioned here, have limitations. Thus, one needs to be cautious when using them for interpreting observations. We should also note that there is still a large gap between observations and models. Therefore, it is still a challenging task to understand the solar magnetic structure and dynamics in a consistent way based on both observations and models. For example, observations of the full Stokes parameters contain much more information than that derived by the Unno--Rachkovsky solution and Milne--Eddington atmosphere model. Without more advanced theoretical models and numerical methods \citep[see,][]{2015Lagg,2016DeLaCruz}, we cannot extract such information precisely. The MHD simulations of solar eruptions should be improved with more physics included and more realistic boundary and initial conditions adopted. Specifically, we should include all necessary physics, such as resistive process, thermal conduction, and radiation, in MHD simulations. The computation domain should be broad enough containing very different physical environments, such as the convection zone, solar atmosphere, and the interplanetary space. Meanwhile, more physical parameters, including magnetic field, velocity field, density, and temperature inferred from multi-wavelength observations, should be assimilated into MHD simulations. This is a long way to go. In a word, just like our knowledge on the earthquake, we expect a continuous reduction of the gap between observations and models of solar eruptions though it may never vanish.


\acknowledgments

The authors thank Prof. X.S. Feng and Prof. W.X. Wan for the invitation to write the review paper, and thank the three anonymous referees for constructive comments that improve the paper. YG, XC, and DMD are supported by NSFC (11533005, 11203014, 11373023, and 11303016) and NKBRSF 2014CB744203.


\bibliographystyle{apj}
\bibliography{bibliography_sci}


\end{document}